\documentclass[11pt]{article}
\usepackage{times}
\usepackage{bbm}
\usepackage{graphicx}
\usepackage{subfigure}

\usepackage{easymat}
\usepackage{chicago}
\usepackage{amsmath}
\usepackage{amssymb}
\usepackage{bbm}
\newcommand{\LB}{\left[\begin{MAT}(r){l}}
\newcommand{\RB}{\\ \end{MAT}\right]}
\newcommand{\D}{\mathbbm{d}}
\newcommand{\F}{\mathbf{f}}
\newcommand{\M}{\mathbf{m}}
\newcommand{\B}{\mathbf{b}}
\newcommand{\W}{\mathbf{w}}
\newcommand{\Bp}{\mathbf{b}^{p}}
\newcommand{\Wp}{\mathbf{w}^{p}}
\newcommand{\Bq}{\mathbf{b}^{q}}
\newcommand{\Wq}{\mathbf{w}^{q}}
\newcommand{\Delfpq}{d\mathbf{f}^{pq}}
\newcommand{\Delmpq}{d\mathbf{m}^{pq}}
\newcommand{\Delfqp}{d\mathbf{f}^{qp}}
\newcommand{\Delmqp}{d\mathbf{m}^{qp}}
\newcommand{\Delbp}{d\mathbf{b}^{p}}
\newcommand{\Delwp}{d\mathbf{w}^{p}}
\newcommand{\Fpq}{\mathbf{f}^{pq}}
\newcommand{\FFpq}{\mathbf{F}^{pq}}
\newcommand{\MMpq}{\mathbf{M}^{pq}}
\newcommand{\Gpq}{\mathbf{g}^{pq}}
\newcommand{\Hpq}{\mathbf{h}^{pq}}
\newcommand{\Fqp}{\mathbf{f}^{qp}}
\newcommand{\Mpq}{\mathbf{m}^{pq}}
\newcommand{\Mqp}{\mathbf{m}^{qp}}
\newcommand{\Delrpq}{d\mathbf{r}^{pq}}
\newcommand{\Rpq}{\mathbf{r}^{pq}}
\newcommand{\Rqp}{\mathbf{r}^{qp}}

\newcommand{\Dtp}{d\boldsymbol{\theta}^{p}}
\newcommand{\Dtq}{d\boldsymbol{\theta}^{q}}
\newcommand{\Dfpq}{\delta\mathbf{f}^{pq}}
\newcommand{\Dmpq}{\delta\mathbf{m}^{pq}}
\newcommand{\Dfqp}{\delta\mathbf{f}^{qp}}
\newcommand{\Dmqp}{\delta\mathbf{m}^{qp}}
\newcommand{\Dbp}{\delta\mathbf{b}^{p}}
\newcommand{\Dwp}{\delta\mathbf{w}^{p}}
\newcommand{\Dut}{[d\mathbf{u}/d\boldsymbol{\theta}]}
\newcommand{\Dbw}{[d\mathbf{b}/d\mathbf{w}]}
\newcommand{\DFfpq}{\mathfrak{d}\mathbf{f}^{pq}}
\newcommand{\DMfpq}{\mathfrak{d}\mathbf{m}^{pq}}
\newcommand{\Drpq}{\delta\mathbf{r}^{pq}}
\newcommand{\Drqp}{\delta\mathbf{r}^{qp}}
\newcommand{\Dsn}{\delta s^{pq\text{, n}}}
\newcommand{\Dst}{\delta s^{pq\text{, t}}}

\newcommand{\Dnpq}{\delta\mathbf{n}^{pq}}
\newcommand{\Dnqp}{\delta\mathbf{n}^{qp}}

\newcommand{\Eone}{\mathbf{e}_{1}}
\newcommand{\Etwo}{\mathbf{e}_{2}}
\newcommand{\Ethree}{\mathbf{e}_{3}}
\newcommand{\Npq}{\mathbf{n}^{pq}}
\newcommand{\Nqp}{\mathbf{n}^{qp}}
\newcommand{\Tpq}{\mathbf{t}^{pq}}
\newcommand{\Tqp}{\mathbf{t}^{qp}}

\newcommand{\Dup}{d\mathbf{u}^{p}}
\newcommand{\Duq}{d\mathbf{u}^{q}}
\newcommand{\Xp}{\mathbf{x}^{p}}
\newcommand{\Xq}{\mathbf{x}^{q}}
\newcommand{\Rhop}{\rho^{p}}
\newcommand{\Rhoq}{\rho^{q}}
\newcommand{\Rhopq}{\Rhop+\Rhoq}
\newcommand{\HH}{\mathbf{H}}
\newcommand{\HHH}{\mathbbm{H}}
\newcommand{\HHHH}{\pmb{\mathcal{H}}}
\newcommand{\Hm}{\mathbf{H}^{\text{m}}}
\newcommand{\Hg}{\mathbf{H}^{\text{g}}}
\newcommand{\Hgone}{\mathbf{H}^{\text{g--1}}}
\newcommand{\Hgtwo}{\mathbf{H}^{\text{g--2}}}
\newcommand{\Hgthree}{\mathbf{H}^{\text{g--3}}}
\newcommand{\DUa}
           {\LB\begin{MAT}(r){l} \Dup\\:\Dtp\\:\Duq\\:\Dtq\\ \end{MAT}\RB}
\newcommand{\DUb}
      {\LB du^p_1\\du^p_2 \\ d\theta^p_3\\: du^q_1\\du^q_2 \\ d\theta^q_3 \RB}
\newcommand{\DMut}{\LB d\mathbf{u}\\:d\boldsymbol{\theta} \RB}
\newcommand{\DMbw}{\LB d\mathbf{b}\\:d\mathbf{w} \RB}
\newcommand{\DDMbw}{\LB \D\mathbf{b}\\:\D\mathbf{w}\RB}
\newcommand{\Dudef}{\delta\mathbf{u}^{pq\text{, def}}}
\newcommand{\Dtdef}{\delta\boldsymbol{\theta}^{pq\text{, def}}}
\newcommand{\Dudefqp}{\delta\mathbf{u}^{qp\text{, def}}}
\newcommand{\Dtdefqp}{\delta\boldsymbol{\theta}^{qp\text{, def}}}
\newcommand{\DuDef}{\delta\mathbf{u}^{\text{def}}}
\newcommand{\DtDef}{\delta\boldsymbol{\theta}^{\text{def}}}
\newcommand{\Kp}{\mathbf{K}^{p}}
\newcommand{\Kq}{\mathbf{K}^{q}}
\newcommand{\Sum}{\sum_{q}}
\newcommand{\Zero}{\mathbf{0}}
\begin{document}
%
%
\thispagestyle{empty}
\title{Stability, Bifurcation, and Softening in Discrete Systems:\\
A Conceptual Approach for Granular Materials}
\author{
Matthew R. Kuhn%
  \thanks{%
  Corresponding author.
  Dept.\ of Civil Engrg.,
  School of Engrg., Univ.\ of Portland, 
  5000 N.\ Willamette Blvd., Portland, OR\ 97203, U.S.A.,
  Tel. 503-943-7361, Fax: 503-943-7316, kuhn@up.edu}
  \ and~%
  Ching S. Chang%
  \thanks{Dept.\ of Civil and Env.\ Engrg., Univ.
  of Massachusetts, Amherst, MA 01002, U.S.A.}
}
\maketitle
\begin{abstract}
Matrix stiffness expressions are derived for the particle movements
in an assembly of rigid granules having compliant contacts.
The derivations include stiffness terms that arise from the particle
shapes at their contacts.  These geometric stiffness terms may
become significant during granular failure.
The geometric stiffness must be added to the 
mechanical stiffnesses of the contacts to produce the complete stiffness.
With frictional contacts, this stiffness expression is 
incrementally nonlinear, having multiple loading branches.
To aid the study of material behavior,
a modified stiffness is derived for 
isolated granular clusters that are considered detached
from the rest of a granular body.
Criteria are presented for bifurcation, instability, and
softening of such isolated and discrete granular sub-regions.
Examples show that instability and softening can result entirely
from the geometric terms in the matrix stiffness.%
\\[1ex]
\end{abstract}
%
%
\section{Introduction} 
\begin{figure}
\centering
\includegraphics[scale=0.82]{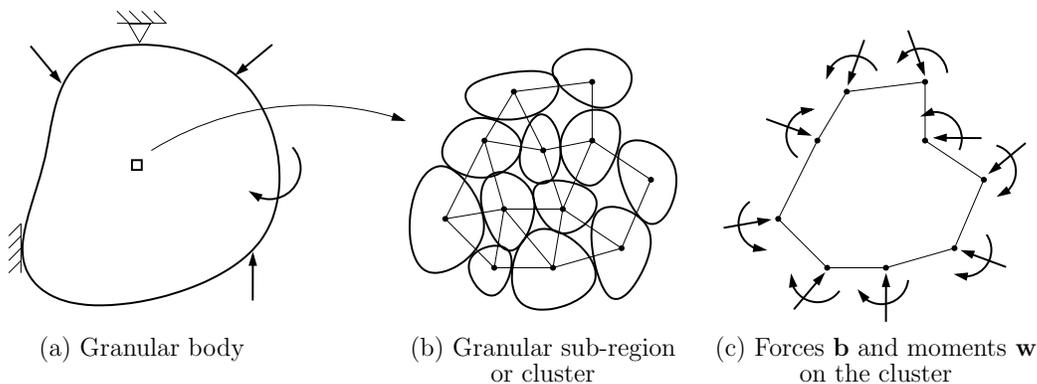}
\caption{Region and sub-region of a granular material.}
\label{fig:bodies}
\end{figure}

\suppressfloats
The paper concerns 
the material behavior of granular media and examines questions
of internal stability, solution uniqueness, 
and softening in these materials.
Granular materials can be viewed as systems of granules
that interact at their points of contact.
The incremental
boundary value problem for a granular system would involve
an entire multi-grain 
body and the prescribed increments (rates) of 
displacements and external forces
(Fig.~\ref{fig:bodies}a).
When viewed as a system of nodes, connections, and
supports, 
the problem resembles conventional problems in structural mechanics.
In an alternative approach,
we could treat the body as a continuum
and investigate uniqueness and stability by evaluating
the material behavior of the entire body or of a representative continuum point
in the manner of \citeN{Hill:1958a}, \citeN{Rice:1976a}, and others.
We suggest that questions of granular behavior can 
be investigated by accepting these materials
as discrete systems, with the intent of
appraising their susceptibility to instability and softening.
The developments in the paper can be applied to the
problem of an entire body and its supports,
although the derivations are primarily directed toward
the problem of \emph{material behavior} within
the body, perhaps the behavior within isolated sub-regions
or representative volume elements (Fig.~\ref{fig:bodies}b).
In either case, the continuum notions of 
stress and deformation are
replaced by discrete contact forces
and particle displacements within
the body or sub-region (Fig.~\ref{fig:bodies}c).
The purpose of this work is to derive the incremental stiffness of a
system of particles---a stiffness that accounts for the
particle shapes---and to provide stability, uniqueness,
and softening criteria for the system.
\par
In Section \ref{sec:mechanical}, we derive the incremental stiffness
matrix for a group of $N$ particles.
The primary contribution of this section is the inclusion
of geometric terms in the derivation, which account for the
shapes of the particles at their contacts.
By including these terms, we show that
the incremental stiffness of a granular material depends, in part,
on the current forces among the particles and not merely
on the contact stiffnesses alone.
The section includes an analysis of possible rigid rotations
of a sub-region when it is considered detached from the rest of
a granular body.
Section~\ref{sec:mechanical} ends with the presentation of
a sample, prototype contact model that can be used in
typical implementations.
In Section~\ref{sec:stability}, we present conditions
for stability, uniqueness, and softening of a granular
sub-region, with particular attention to the
incrementally nonlinear behavior of contacts within the sub-region.
Section~\ref{sec:example} presents examples of two-particle 
and four-particle systems,
and we end by discussing implications of this work
and possible future directions.
A list of notation is given in Appendix~\ref{sec:notation}, and
some derivations are placed in Appendices~\ref{app:derive}--\ref{app:eigen}.
\section{Stiffness of a granular region}\label{sec:mechanical}
We consider the incremental motions and stiffness 
of an assembly or cluster of particles
(Fig.~\ref{fig:bodies}b).
Particle positions, contact forces, and loading history are
assumed known at the current time $t$, 
insofar as they affect the current incremental contact stiffnesses.
We address the incremental (or rate) problem in which
certain infinitesimal particle motions 
and external force increments are prescribed,
and we seek the remaining, unknown motion and force increments.
The particles are assumed to be smooth and durable, with no
particle breaking,
and particles interact solely at their contacts (i.e., no long-range
inter-particle forces).
The particles are also assumed to be rigid except at their
compliant contacts, where the traction between a pair
of particles is treated as a point force that depends
on the relative motions of the two particles.
For example,
this assumption would be consistent with Hertz-type contact
models in which changes in force are produced by the relative
approach of two particles.
This compliant contact viewpoint
differs, however, from ``hard contact'' models that enforce 
unilateral force and displacement constraints \cite{Moreau:2004a}.
Finally,
we assume slow deformations and rate-independent contact behavior.
\par
\begin{figure}
\centering
\includegraphics[scale=0.90]{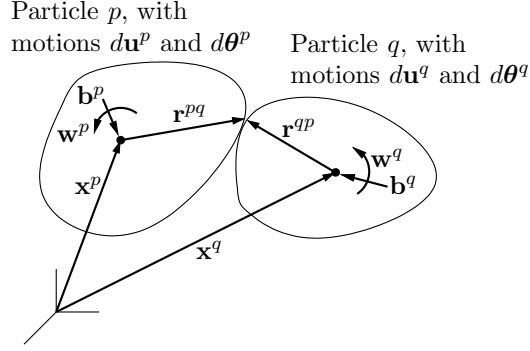}
\caption{Two particles in contact.}
\label{fig:particles}
\end{figure}

With these assumptions,
particle motions are governed by the mechanics of
rigid bodies with compliant contacts:
particle motions produce contact deformations; contact deformations
produce contact forces; and the forces on each particle
must be in equilibrium.
In this section, we derive the stiffness equation 
for a three-dimensional group (or cluster) of $N$ particles in the form
\begin{equation}\label{eq:H}
\LB \mathbf{H} \RB_{6N \times 6N}
\DMut_{6N \times 1}
= \DMbw_{6N \times 1}
\end{equation}
where $[\mathbf{H}]$ is the incremental stiffness matrix, 
vector $[d\mathbf{u}/ d\boldsymbol{\theta}]$ 
contains three incremental displacements
and three incremental rotations for each of the $N$ particles,
and vector $[d\mathbf{b}/ d\mathbf{w}]$ contains the
six infinitesimal
increments of external force and moment applied to each of the $N$
particles (Fig.~\ref{fig:particles}).
The derivation allows for both contact forces and contact moments,
as well as for both
external body forces $d\mathbf{b}$ and external body moments $d\mathbf{w}$.
These external forces may embody the influence
of surrounding particles on the cluster,
and the paper is primarily directed toward problems in which 
the increments $\Dbw$ are prescribed and the displacements
$\Dut$ must be solved.
In the derivations, we include
all stiffness terms of order $(du)^{1}$ but
exclude terms of higher order.
Even so, Eq.~(\ref{eq:H}) may lead to instabilities,
just as a small strain--finite rotation approach can uncover
instabilities in continuous systems.
The results show that the cluster stiffness 
does not exclusively depend upon the
stiffnesses of the contacts (i.e., on the ``contact springs''); 
instead, the incremental stiffness also 
includes geometric contributions that
depend on the shapes of particles
at their contacts and on the current, accumulated contact forces.
%
\par
The stiffness matrix $[\mathbf{H}]$ can be assembled in a conventional
manner from the stiffness matrices of the assembly's elemental units---the
individual contacts between particle pairs---and this
section is primarily concerned with deriving the incremental
stiffness of a single pair of particles.
Consider two representative
particles, $p$ and $q$, that are in contact (Fig.~\ref{fig:particles}).
The incremental stiffness contributed
by this one contact can be expressed in matrix form as
\begin{equation}\label{eq:piece}
\left[
\begin{MAT}(e){l:l}
\mathbf{H}^{p\text{--}p} & \mathbf{H}^{p\text{--}q} \\:
\mathbf{H}^{q\text{--}p} & \mathbf{H}^{q\text{--}q} \\
\end{MAT}
\right]_{12 \times 12}
\LB d\mathbf{u}^{p}\\ \Dtp\\: d\mathbf{u}^{q}\\ \Dtq \RB_{12 \times 1}
=
\LB d\mathbf{b}^{p,\,pq}\\
    d\mathbf{w}^{p,\,pq}\\:
    d\mathbf{b}^{q,\,qp}\\
    d\mathbf{w}^{q,\,qp}\RB_{12 \times 1}
\end{equation}
where $\Dup$, $\Duq$, $\Dtp$, and $\Dtq$ are the translations and rotations
of $p$ and $q$.
Equation~(\ref{eq:piece}) expresses the effect that the single contact
between $p$ and $q$ will have upon the equilibrium of the two particles.
The external force increments on the right of Eq.~(\ref{eq:piece})
must be combined with the forces that are implied by 
the other contacts in an assembly or cluster.
The stiffness matrices of all $M$ contacts within the cluster
can be assembled in the usual manner into
a global matrix---the matrix $[\mathbf{H}]$ of Eq.~(\ref{eq:H}).
The matrix assembly process has been described
elsewhere in the context of the 
finite element method (FEM), discontinuous deformation
analysis (DDA), and the granular element method (GEM)
(see \citeNP{Bathe:1976a}, \citeNP{Shi:1993a}, and \shortciteNP{Kaneko:2003a},
respectively).
In the current work,
we do not consider boundary constraints 
(prescribed displacements) on the cluster, and this absence
will, of course, leave $[\mathbf{H}]$ singular,
with rigid-body modes of motion.
The possibility of such rigid modes will affect our assessment
of stability, a matter that we consider in Section~\ref{sec:rigid_rotation}.
\subsection{Objective incremental vectors}\label{eq:objectivity}
In deriving Eqs.~(\ref{eq:H}) and~(\ref{eq:piece}),
we preferentially use \emph{objective} incremental vectors,
since the response of a granular sub-region or contact should be
independent of the observer, even if the observer is moving
(\citeNP{Truesdell:1960a}, \S293).
An incremental vector is objective if it is assigned the same
measure by two moving observers who briefly share the
same frame at time $t$ but then rotate relative to one
another during the interval of $t$ to $t+dt$.
The increment $d\mathbf{y}$ between the initial and final
vectors $\mathbf{y}^{t}$ and $\mathbf{y}^{t+dt}$,
\begin{equation}
d\mathbf{y} = \mathbf{y}^{t+dt} - \mathbf{y}^{t}\;,
\end{equation}
is not objective, since an observer who rotates with $\mathbf{y}$
would observe a different $d\mathbf{y}$ than would a
stationary observer.
The discrepancy is corrected, of course, when the two observers
independently measure some other angular change $d\boldsymbol{\theta}$
that occurs during $dt$.
For example, if $d\boldsymbol{\theta}$ is the observed
rotation of the direction of $\mathbf{y}^{t+dt}$ relative
to $\mathbf{y}^{t}$, then the corotated force
\begin{equation}
\mathbf{y}^{t\text{, corotated}} = \mathbf{y}^{t} + 
d\boldsymbol{\theta}\times\mathbf{y}^{t}
\end{equation}
can be subtracted from $\mathbf{y}^{t+dt}$ to compute an increment
$\Delta \mathbf{y}$ that would be assigned the same measure by both observers:
\begin{equation}\label{eq:Deltav}
\Delta \mathbf{y} = \mathbf{y}^{t+dt} - \mathbf{y}^{t\text{, corotated}}
= d\mathbf{y} - d\boldsymbol{\theta}\times\mathbf{y}^{t}\;.
\end{equation}
The increment $\Delta \mathbf{y}$ is objective.
Other objective increments can be extracted by referencing other
rotations $d\boldsymbol{\theta}$.
\par
In the paper, 
we use four types of infinitesimal increments---designated by the
symbols $d$, $\delta$, $\mathfrak{d}$, and
$\mathbbm{d}$---with the following
distinctions:
\begin{itemize}
\item
``$d$'' increments are those seen by a distant (and possibly moving)
observer and are not objective.
\item
``$\delta$'' increments are those viewed by an
observer attached to (and moving with) a single particle
(the angle $ d\boldsymbol{\theta}$ in Eq.~\ref{eq:Deltav} is
taken as the particle rotation).
These increments are objective.
\item
``$\mathfrak{d}$'' increments are also objective but are tied to the
local material characteristics of two particles at their contact
(the angle $ d\boldsymbol{\theta}$ in Eq.~\ref{eq:Deltav} is
taken as the rotation of the contact frame as the particles rotate or twirl
across each other).
\item
``$\mathbbm{d}$'' increments are objective projections of force and
displacement onto certain objective subspaces
(Section~\ref{sec:rigid_rotation}, where
the angle $ d\boldsymbol{\theta}$ in Eq.~\ref{eq:Deltav} is taken as the
average rotation of a particle cluster).
\end{itemize}
\subsection{First geometric stiffness}\label{sec:firstgeometry}
The current contact forces $\F$ and the current contact
moments $\M$ on a single particle $p$ are assumed to be known
\emph{a prior} and to be in equilibrium
with the external force and moment:
\begin{equation}\label{eq:equil1}
-\Sum\Fpq = \Bp \;, \quad -\Sum\left(\Rpq\times\Fpq + \Mpq \right)= \Wp \;,
\end{equation}
where the sums are for all particles ``$q$'' that are in
contact with $p$,
and $\Bp$ and $\Wp$ are the current external body force and body moment
that act upon $p$ through the current position $\mathbf{x}^{p}$ of
its pre-assigned (material) reference point (Fig.~\ref{fig:particles}).
The internal contact force $\Fpq$ and contact moment $\Mpq$
act upon particle $p$ at its contact point with $q$, and
the radial vector $\Rpq$ is directed from the
reference point $\mathbf{x}^{p}$ of $p$ to the contact point with $q$.
In contrast, $\Fqp$ and $\Mqp$ act upon particle $q$, and 
$\Rqp$ is directed from 
the point $\mathbf{x}^{q}$ in particle $q$.
\par
The incremental forms of Eqs.~(\ref{eq:equil1}$_{1}$) 
and~(\ref{eq:equil1}$_{2}$) are
\begin{equation}\label{eq:equil2}
-\Sum\Delfpq = d\Bp \;, \quad 
-\Sum(\Delrpq\times\Fpq + \Rpq\times\Delfpq +\Delmpq) = d\Wp \;,
\end{equation}
where we account for changes $\Delrpq$ in the radii as well
as changes $\Delfpq$ and $\Delmpq$ in the contact forces.
As such, we pursue a second-order theory which accounts for
equilibrium in the deflected shape.
An infinitesimal ``$d$'' increment is one seen by a distant, possibly
moving, observer.
None of the incremental ``$d$'' vectors 
in Eq.~(\ref{eq:equil2}) are objective, 
but we can identify
an objective ``$\delta$'' part of each increment:
\begin{align}
\label{eq:dr}
\Delrpq &= \Drpq + \Dtp\times\Rpq
\\
\label{eq:df}
\Delfpq &= \Dfpq + \Dtp\times\Fpq
\\
\label{eq:dm}
\Delmpq &= \Dmpq + \Dtp\times\Mpq
\\
\label{eq:db}
\Delbp &= \Dbp + \Dtp\times\Bp
\\
\label{eq:dw}
\Delwp &= \Dwp + \Dtp\times\Wp
\end{align}
where $\Dtp$ is the incremental rotation of particle $p$.
The objective ``$\delta$'' increments 
are those that would be viewed by an observer attached to
and moving with the particle $p$; whereas,
the cross products in Eqs.~(\ref{eq:dr})--(\ref{eq:dw}) are the increments
that would be seen by a stationary observer when viewing
a vector (say, a follower force $\Bp$) that happens to be 
rotating in unison with the particle.
Although the force increments on particles $p$ and $q$ are self-equilibrating,
with $\Delfpq=-\Delfqp$ and $\Delmpq=-\Delmqp$,
the corotating increments $\Dfpq$ and $\Dmpq$ are not 
necessarily equal to the negatives of their
counterparts, $-\Dfqp$ and $-\Dmqp$, since the
``$pq$'' and ``$qp$'' increments are viewed by different observers.
\par
The equilibrium Eqs.~(\ref{eq:equil2}$_1$) and~(\ref{eq:equil2}$_2$)
can also be expressed in terms of objective ``$\delta$'' increments,
\begin{align}
\label{eq:equil3}
&-\Sum\Dfpq = \Dbp \\
\label{eq:equil4}
&-\Sum\left(\Drpq\times\Fpq + \Rpq\times\Dfpq + \Dmpq\right) = \Dwp \;,
\end{align}
as derived in Appendix~\ref{app:derive}.
%
As expected, incremental equilibrium is an objective relationship,
independent of the observer, and expressible in terms of objective
quantities.
\par
An infinitesimal change in the radial contact position,
$\delta\Rpq$ in Eq.~(\ref{eq:equil4}),
alters the moment equilibrium of particle $p$.
This effect is related to similar geometric effects in structural
mechanics, such as buckling and ``p-delta'' phenomena that arise
from the flexing or swaying of columns and frames.
The increment $\delta\Rpq$ is objective and
can be separated into normal and tangential parts, which are
both amenable to kinematic/geometric analysis:
\begin{equation}\label{eq:Drpq}
\Drpq = \Dsn\Npq + \Dst\Tpq\;.
\end{equation}
In this equation, $\Npq$ and $\Tpq$ are unit vectors
in directions normal and tangential to $p$ at its contact with $q$, 
and $\Dsn$ and $\Dst$ are the associated displacement magnitudes.
Note that $\Npq=-\Nqp$,
but the increments $\Drpq$ and $\Dst\Tpq$
might not equal the negatives of their counterparts
$\Drqp$ and $\delta s^{qp\text{, t}}\Tqp$,
since the latter are viewed by an observer attached to $q$.
\par
For a compliant contact, the normal displacement 
$\Dsn\Npq$ can be taken as 
the average incremental indentation of the two particles:
\begin{equation}\label{eq:dsn}
\Dsn\Npq = \frac{1}{2}\left( \Dudef\cdot\Npq\right)\Npq\;,
\end{equation}
where the objective vector $\Dudef$ is the 
translation of $p$ relative to $q$ near their contact, 
\begin{equation}\label{eq:dudef}
\Dudef = \Duq -\Dup + \left( \Dtq\times\Rqp-\Dtp\times\Rpq\right)\;,
\end{equation}
with $\Dudef=-\Dudefqp$.
\par
The displacement $\Dst\Tpq$ is the tangential movement of the
contact point, as viewed by an observer attached to $p$,
a movement that is produced by a combination of sliding and rolling
motions, described by \citeN{Kuhn:2004b},
%
\begin{multline}\label{eq:rolling}
\Dst\Tpq = \\
          -\left( \Kp + \Kq \right)^{-1} \cdot
           \left[
                 \Dtdef\times\Npq -
                 \Kq\cdot\left( \Dudef - (\Dudef\cdot\Npq)\Npq\right)
           \right]
\end{multline}
where the objective rotational contact deformation $\Dtdef$ is
defined as
\begin{equation}\label{eq:dtdef}
\Dtdef = \Dtq - \Dtp \;,
\end{equation}
with $\Dtdef=-\Dtdefqp$.
Tensors $\Kp$ and $\Kq$ are the surface curvatures of particles $p$ and $q$
at their contact, with negative curvatures (eigenvalues) associated
with convex particles.
Both positive and negative curvatures are allowed in the paper, 
provided that particle surfaces are sufficiently smooth---having
continuous curvatures at the contacts points.
We note, however, that a pseudo-inverse should be used in place of 
$(\Kp + \Kq)^{-1}$,
so that the rolling displacement vector $\Dst\Tpq$
is projected onto the tangent plane \cite{Kuhn:2004b}.
\par
Both of the increments 
$\Dsn\Npq$ and $\Dst\Tpq$ are objective, since both
are linear combinations of the objective vectors
$\Dudef$ and $\Dtdef$.
In presenting Eqs.~(\ref{eq:dsn}) and~(\ref{eq:rolling}),
we have intensionally ignored changes in the curvatures
that are produced by particle deformations, since such changes
would produce force increments of an order higher than $(du)^{1}$.
\par
Having developed expressions for the 
$\Drpq$ in Eq.~(\ref{eq:equil4}), 
we anticipate, however, that the contribution of the normal displacement
$\Dsn\Npq\times\Fpq$ 
is likely small, and its effect is probably inconsequential
when compared with the product $\Rpq\times\Dfpq$
in Eq.~(\ref{eq:equil4}).
On the other hand,
the tangential terms $\Dst\Tpq\times\Fpq$ will likely
become significant, perhaps dominant, at larger strains, since particle
rolling becomes a prevailing mechanism 
during granular failure \cite{Kuhn:2004k}.
\par
\begin{figure}%
\centering%
\includegraphics[scale=0.80]{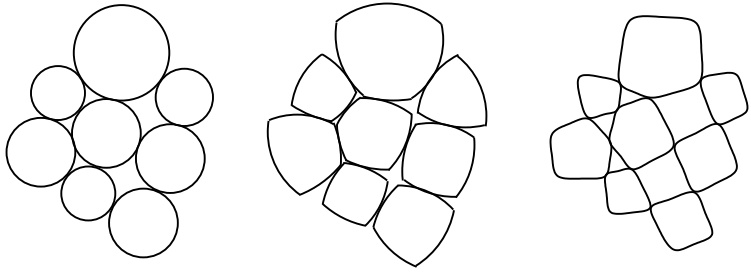}%
\caption{Three clusters with the same topological arrangement, but
different particle curvatures at their contacts.}%
\label{fig:Geometry}%
\end{figure}

Equation~(\ref{eq:equil4}) includes the effects of the
$\Drpq$ increments on the equilibrium of the
single particle $p$, and the similar effects upon all $N$ particles
can be collected into a matrix form as
\begin{equation} \label{eq:Hg1}
-\Sum\Drpq\times\Fpq
\ \rightsquigarrow\ %
\LB \Hgone \RB_{6N \times 6N}
\LB d\mathbf{u}\\: d\boldsymbol{\theta} \RB_{6N \times 1}\;.
\end{equation}
where matrix $[\Hgone ]$ 
is computed with Eqs.~(\ref{eq:equil4})--(\ref{eq:dtdef}).
When constructing the matrix $[ \Hgone ]$,
one must include the separate contributions of
$\Drpq\times\Fpq$ and $\Drqp\times\Fqp$,
which pertain to the equilibrium of particles $p$ and $q$, respectively.
The symbol ``$\rightsquigarrow$'' connotes a matrix assembly
process that collects multiple equilibrium relations
in the form of Eq.~(\ref{eq:piece}) for all $N$ particles.
The six equilibrium equations~(\ref{eq:equil3}) and~(\ref{eq:equil4}),
which apply to any single particle,
can be gathered into the $6N$ equilibrium equations,
\begin{equation}\label{eq:Matrix2}
  \LB \Hgone \RB_{6N \times 6N}
  \DMut_{6N \times 1}
- \LB \mathbf{A}_{1} \RB_{6N \times 2(6M)}
  \LB \delta\mathbf{f}\\: \delta\mathbf{m} \RB_{2(6M) \times 1}
= \LB \delta\mathbf{b}\\: \delta\mathbf{w} \RB_{6N \times 1} \;,
\end{equation}
by collecting the contact force increments,
$\delta\mathbf{f}^{(\cdot )}$ and $\delta\mathbf{m}^{(\cdot )}$,
of all $M$ contacts.
The first matrix product,
$[\Hgone ] \Dut$,
corresponds to the quantities
$\Drpq\times\Fpq$ in Eqs.~(\ref{eq:equil4})--(\ref{eq:Hg1});
the second product 
$[\mathbf{A}_{1}][\delta\mathbf{f}/\delta\mathbf{m}]$
corresponds to the 
$\Dfpq$ and $\Dmpq$ terms in Eqs.~(\ref{eq:equil3}) and~(\ref{eq:equil4}).
These latter terms will soon be investigated.
When assembling the contact forces and moments
into Eq.~(\ref{eq:Matrix2}),
we use a less conventional approach:  the
contact forces $\Dfpq$ and $\Dfqp$ are treated as distinct objects,
since $\Dfpq$ and $\Dmpq$ are not usually equal to $-\Dfqp$ and $-\Dmqp$.
This distinction 
leads to a total of $2(6M)$ contact force/moment components among
the $M$ contacts.  
The statics matrix $[\mathbf{A}_{1}]$
combines these contact forces and moments, as with the 
$\Dfpq$ and $\Dmpq$ sums of
Eqs.~(\ref{eq:equil3}) and~(\ref{eq:equil4}).
Although it may be impossible to entirely separate geometric and
mechanical effects, 
the $[\Hgone ]$ product in~(\ref{eq:Matrix2})
originates from the geometric, surface shapes of the particles
and from the current contact forces $\Fpq$ and $\Mpq$.
The matrix $[\Hgone ]$ would differ for the three 
clusters in Fig.~\ref{fig:Geometry}
and would partially account for any differences in
their incremental responses.
Other geometric effects will arise
from the $[\delta\mathbf{f}/\delta\mathbf{m}]$ vector of 
Eq.~(\ref{eq:Matrix2}), which will now be discussed.%
%
%
\subsection{Mechanical stiffness; 
            second and third geometric stiffnesses}\label{sec:submechanical}
To achieve the form of Eq.~(\ref{eq:H}),
the product 
$[\mathbf{A}_{1}][\delta\mathbf{f}/\delta\mathbf{m}]$
in Eq.~(\ref{eq:Matrix2}) must be expressed in terms of the $6N$
particle movements $\Dut$.
The increments of a single contact's force and moment
will depend upon the contact deformations
of the two particles and also upon any change in the orientation
of their contact plane.
The increments of force and moment can be derived in terms of
either the ``$\delta$'' or ``$d$'' increments.
Using the simpler ``$d$'' increments, as viewed by a distant observer,
\begin{align}\label{eq:delfpq}
\Delfpq &= \DFfpq + \Fpq\times\left( d\mathbf{n}^{pq}\times\Npq\right)
  - \frac{1}{2}\left[ \left( \Dtp + \Dtq\right)\cdot\Npq\right]\Fpq\times\Npq
\\
\label{eq:delmpq}
\Delmpq &= \DMfpq + \Mpq\times\left( d\mathbf{n}^{pq}\times\Npq\right)
 - \frac{1}{2}\left[ \left( \Dtp + \Dtq\right)\cdot\Npq\right]\Mpq\times\Npq\;.
\end{align}
The increments $\DFfpq$ and $\DMfpq$ are the objective changes
in contact force and moment produced solely by material deformations
of the two particles near their contact.
These increments depend upon the objective deformation
vectors $\Dudef$ and $\Dtdef$, 
and the possible nature of this dependence will be discussed later.
The terms $\Fpq\times\left( d\mathbf{n}^{pq}\times\Npq\right)$
and $\Mpq\times\left( d\mathbf{n}^{pq}\times\Npq\right)$ are the 
force increments
produced by a rotation (tilting) of the contact plane, as seen
by a distant ``$d$'' observer.
These terms are typically computed in DEM algorithms in the
manner of \citeN{Lin:1997a}
and \shortciteN{Vuquoc:2000a}.
The final, subtracted terms in Eqs.~(\ref{eq:delfpq}) and~(\ref{eq:delmpq})
are not yet encountered in the DEM literature and
are produced by a rigid-body twirling of the particle pair.
That is, a rigid twirling of two particles, with $\Dtp=\Dtq=d\theta\Npq$,
will leave the normal direction $\Npq$ unchanged but
will cause the tangential contact force to rotate 
with the particles in the plane of their contact.
(Alternatively,
an apparent rotation of force would be seen in a stationary pair of
particles when viewed by a distant observer who is
twirling about the direction $\Npq$.)
The rotations $\Dtp$ and $\Dtq$ are assigned equal weight in
Eq.~(\ref{eq:delfpq}), so that $\Delfpq$ 
will equal $-\Delfqp$ when $p$ and $q$ are interchanged
(see \citeNP{Bagi:2005a}).
\par
Equation~(\ref{eq:delfpq}) can also be written in terms of the
corotated, objective ``$\delta$'' vectors, as required
in Eqs.~(\ref{eq:equil3}) and~(\ref{eq:equil4}):
\begin{equation}\label{eq:Dfpq}
\Dfpq = \DFfpq + \Fpq\times\left( \Dnpq\times\Npq\right)
         - \frac{1}{2}\left(\Dtdef\cdot\Npq\right)\Fpq\times\Npq\;,
\end{equation}
which is derived in Appendix~\ref{app:derive}.
In Eqs.~(\ref{eq:delfpq}) and~(\ref{eq:Dfpq}), 
the total change in the contact normal, $d\mathbf{n}^{pq}$, is the
sum of two parts,
\begin{equation}\label{eq:dn}
d\mathbf{n}^{pq} = \Dnpq + \Dtp\times\Npq\;,
\end{equation}
in the manner of Eqs.~(\ref{eq:dr})--(\ref{eq:dw}),
and these two parts will be discussed later.
As expected, the objective, corotated increment
$\Dfpq$ in Eq.~(\ref{eq:Dfpq}) depends solely 
on other objective quantities---those vectors on the right side
of Eq.~(\ref{eq:Dfpq}).
Likewise, the corotated moment increment is
\begin{equation}
\label{eq:Dmpq}
\Dmpq = \DMfpq + \Mpq\times\left( \Dnpq\times\Npq\right)
         - \frac{1}{2}\left(\Dtdef\cdot\Npq\right)\Mpq\times\Npq\;.
\end{equation}
The increments $\DFfpq$ and $\DMfpq$
depend upon the infinitesimal contact deformations $\Dudef$ and $\Dtdef$,
but the other increments 
depend upon the local shapes of the two particles at their contact
and upon the accumulated, current contact force $\Fpq$ and $\Mpq$.
The $\Dnpq$ terms in Eqs.~(\ref{eq:Dfpq})
and~(\ref{eq:Dmpq}) are likely insignificant at
small strains, but they may become dominant when the
material is failing \cite{Kuhn:2003h}.
\par
Returning to Eq.~(\ref{eq:dn}),
the second term on its right is the change
in the normal $\mathbf{n}^{pq}$ that would be produced 
by a rigid rotation of the particle pair that occurs with no
change in the contact point on the surface of particle $p$.
This term is not objective.
The objective
increment $\Dnpq$ in Eq.~(\ref{eq:dn}) is the change in the normal
that results from a relocation of the contact point on particle $p$, as viewed
by an observer attached to (and rotating with) $p$.
We note, however, that
an observer attached to $q$ will likely view a different reorientation
$\Dnqp$ of its contact point with $p$.
The increment $\Dnpq$ depends upon the
curvature of particle $p$ and is (see \citeNP{Kuhn:2004b})
\begin{equation}\label{eq:Dnpq}
\Dnpq = -\Kp\cdot(\Dst\Tpq)\;,
\end{equation}
where the contact displacement $\Dst\Tpq$ is given in Eq.~(\ref{eq:rolling}).
The force increments 
in the final two terms of Eqs.~(\ref{eq:Dfpq}) and~(\ref{eq:Dmpq})
are collected into a matrix form by 
applying Eqs.~(\ref{eq:Dnpq}) and~(\ref{eq:rolling}) to all
$M$ contacts:
\begin{multline}\label{eq:A3}
\left[
\begin{MAT}(r){l}
\Fpq\times(\Dnpq\times\Npq)- (1/2)\left(\Dtdef\cdot\Npq\right)\Fpq\times\Npq\\:
\Mpq\times(\Dnpq\times\Npq)- (1/2)\left(\Dtdef\cdot\Npq\right)\Mpq\times\Npq\\:
\Fqp\times(\Dnqp\times\Nqp)- (1/2)\left(\Dtdef\cdot\Nqp\right)\Fqp\times\Nqp\\:
\Mqp\times(\Dnqp\times\Nqp)- (1/2)\left(\Dtdef\cdot\Nqp\right)\Mqp\times\Nqp\\
\end{MAT}
\right]_{2(6M)\times 1}
\\
\rightsquigarrow\ %
\LB \mathbf{A}_{2} \RB_{2(6M) \times 6N}
\LB d\mathbf{u}\\: d\boldsymbol{\theta} \RB_{6N \times 1}\;,
\end{multline}
\par
We now consider the remaining terms, $\DFfpq$ and $\DMfpq$,
that appear in Eqs.~(\ref{eq:Dfpq}) and~(\ref{eq:Dmpq}).
Unique injective mappings
are assumed from the full $\mathbb{R}^{6}$ 
space of incremental contact deformations,
$\Dudef$ and $\Dtdef$, into the possibly smaller space of incremental
contact force and moment,
$\DFfpq$ and $\DMfpq$.
We also assume that the particles are rigid except
at their compliant contacts.
For such contact between two particles, 
any objective increment of contact force or moment,
such as $\DFfpq$ or $\DMfpq$, must depend on the
objective, relative increments $\Dudef$ and $\Dtdef$ of their
movements \cite{Kuhn:2005c}.
The assumption of a unique mapping 
$[\Dudef / \Dtdef] \rightarrow [\DFfpq / \DMfpq]$
excludes Signorini models of contact behavior.
Finally, we assume that the mapping is homogeneous of degree one
in both $\Dudef$ and $\Dtdef$, perhaps in the restricted form
%
%
%
\begin{align}\label{eq:constitutive}
\DFfpq
&=
\mathbf{F}^{pq}\left( \frac{\Dudef}{|\Dudef |}\,,\, \Fpq\right)\cdot\Dudef
\\
\label{eq:constitutiveM}
\DMfpq
&=
\mathbf{M}^{pq}\left( \frac{\Dtdef}{|\Dtdef |}\,,\, \Mpq\right)\cdot\Dtdef
\;.
\end{align}
where we introduce the 
contact stiffness tensor functions
$\mathbf{F}^{pq}$ and $\mathbf{M}^{pq}$,
noting that $\mathbf{F}^{pq}=-\mathbf{F}^{qp}$ 
and $\mathbf{M}^{pq}=-\mathbf{M}^{qp}$.
We could also choose more
general forms of contact behavior than those in
Eq.~(\ref{eq:constitutive}) and~(\ref{eq:constitutiveM}).
In these equations,
we have excluded viscous effects 
(see \shortciteNP{Poschel:2001a}), 
but we allow the incremental response to
depend on the current contact force $\Fpq$, as would apply with
frictional contacts.
The constitutive forms~(\ref{eq:constitutive})
and~(\ref{eq:constitutiveM}) depend upon the directions
of the deformations $\Dudef$ and $\Dtdef$ and are,
at best, incrementally nonlinear, as would be expected for 
frictional contacts.
For general Mindlin-Cattaneo contacts, the form would additionally need to
include the history of the contact force.
We also note that in Eqs.~(\ref{eq:constitutive}) 
and~(\ref{eq:constitutiveM}),
a contact's force and moment are uncoupled from each other and
are also uncoupled from the forces and moments at the other
contacts of the same particle, although the latter condition may not
be suitable for very soft particles.
The forms in Eqs.~(\ref{eq:constitutive}) and~(\ref{eq:constitutiveM}) 
would also 
not be appropriate for capturing the effects
of rolling friction, in which $\DFfpq$ and $\DMfpq$
depend on a combination of the translational
and rotational deformations, $\Dudef$ and $\Dtdef$ 
\shortcite{Iwashita:1998a,Vuquoc:2000a}.
Section~\ref{sec:epstiffness} recounts a specific example of the behavior
in Eq.~(\ref{eq:constitutive}).
\par
The general stiffness relations in
Eqs.~(\ref{eq:constitutive}) and~(\ref{eq:constitutiveM})
are collected
for all $M$ contacts into the matrix form
\begin{equation}\label{eq:constitMatrix}
\LB \mathfrak{d}\mathbf{f}\\: \mathfrak{d}\mathbf{m} \RB_{2(6M) \times 1}
=
\left[
\begin{MAT}(r)[0pt,2.5em,2em]{c}
\mathbf{F}\\: 
\mathbf{M}\\  
\end{MAT}
\right]_{2(6M) \times 6M}
\LB \DuDef\\: \DtDef \RB_{6M \times 1}\;,
\end{equation}
recognizing that the contents of matrix $[\mathbf{F}/\mathbf{M}]$
may depend upon the current contact forces, $\mathbf{f}^{pq}$ and
$\mathbf{m}^{pq}$,
and on the directions of the incremental contact deformations, $\Dudef$
and $\Dtdef$.
That is, the mapping from $[\Dudef / \Dtdef ]$
to $[\mathfrak{d}\mathbf{f}/ \mathfrak{d}\mathbf{m}]$
may be incrementally nonlinear in a manner explored 
in Sections~\ref{sec:epstiffness} and~\ref{sec:stability}.
To be consistent with Eqs.~(\ref{eq:Matrix2}) and~(\ref{eq:A3}),
we treat the forces $\mathfrak{d}\mathbf{f}^{pq}$
and $\mathfrak{d}\mathbf{m}^{pq}$
as being distinct from $\mathfrak{d}\mathbf{f}^{qp}$ 
and $\mathfrak{d}\mathbf{m}^{qp}$,
even though
$\mathfrak{d}\mathbf{f}^{pq}=-\mathfrak{d}\mathbf{f}^{qp}$,
$\mathfrak{d}\mathbf{m}^{pq}=-\mathfrak{d}\mathbf{m}^{qp}$,
$\mathbf{F}^{pq}=-\mathbf{F}^{qp}$,
and $\mathbf{M}^{pq}=-\mathbf{M}^{qp}$.
\par
The contact deformations $\Dudef$ 
and $\Dtdef$
in Eqs.~(\ref{eq:constitutive})--(\ref{eq:constitMatrix})
depend upon the motions of the two particles $p$ and $q$.
These kinematic relationships are supplied by
Eqs.~(\ref{eq:dudef}) and~(\ref{eq:dtdef}),
which can be collected in a matrix form as
\begin{equation}\label{eq:kinematics}
  \LB \DuDef\\: \DtDef \RB_{6M \times 1}
= \LB \mathbf{B} \RB_{6M \times 6N}
  \DMut_{6N \times 1}
\end{equation}
for all $N$ particles and their $M$ contacts.
Matrix $[\mathbf{B}]$ is the kinematics matrix.
%
%
\par
Equations~(\ref{eq:Dfpq}), (\ref{eq:Dmpq}), (\ref{eq:A3}),
(\ref{eq:constitMatrix}) and~(\ref{eq:kinematics})
are substituted into Eq.~(\ref{eq:Matrix2})
to arrive at a matrix equation for all particle
motions within a granular assembly:
\begin{equation}\label{eq:prefinal}
\left(
\LB \Hgone \RB 
+ \LB \mathbf{H}^{\text{g--2}} \RB 
+
\LB \Hm \RB
\right)
\DMut
= \LB \delta\mathbf{b}\\: \delta\mathbf{w} \RB_{6N \times 1} \;,
\end{equation}
where the ``mechanical'' stiffness $[ \Hm ]$ is
\begin{equation} \label{eq:Hm}
\LB \Hm \RB_{6N\times 6N} =
-\LB \mathbf{A}_{1} \RB_{6N\times 2(6M)}
\left[
\begin{MAT}(r)[0pt,2em,2em]{c}
\mathbf{F}\\: 
\mathbf{M}\\  
\end{MAT}
\right]_{2(6M)\times 6M}
\LB \mathbf{B} \RB_{6M\times 6N}
\end{equation}
and the second geometric stiffness $[\Hgtwo ]$ is 
\begin{equation}\label{eq:equilibrium4}
\LB \mathbf{H}^{\text{g--2}} \RB_{6N\times 6N}
=
-\LB \mathbf{A}_{1} \RB_{6N\times 2(6M)}
 \LB \mathbf{A}_{2} \RB_{2(6M)\times 6N} \;.
\end{equation}
This geometric stiffness accounts for the rotations of
contact forces that accompany the rolling and twirling of particle pairs.
The stiffness 
$[ \Hm ]$
in Eq.~(\ref{eq:Hm})
is the conventional mechanical stiffness matrix
for a system of $N$ nodes that interact through $M$ connections,
but in a granular system, the connections are through contacts whose
positions and orientations are altered by the particle movements---even
infinitesimal movements.
The geometric alterations are captured, in part, 
with the matrices $[ \Hgone ]$
and $[\mathbf{H}^{\text{g--2}}]$. 
A third alteration is also required.
\par
To attain the desired form of Eq.~(\ref{eq:H}),
the corotating forces $\delta\mathbf{b}$ and
$\delta\mathbf{w}$ must be converted into the conventional increments
$d\B$ and $d\W$.
In view of Eqs.~(\ref{eq:db}) and~(\ref{eq:dw}),
\begin{equation}\label{eq:A4}
\LB d\mathbf{b}\\: d\mathbf{w} \RB_{6N \times 1}
= \LB \delta\mathbf{b}\\: \delta\mathbf{w} \RB_{6N \times 1}
+ \LB \mathbf{H}^{\text{g--3}} \RB_{6N\times 6N}
  \LB d\mathbf{u}\\: d\boldsymbol{\theta} \RB_{6N \times 1}\;,
\end{equation}
where the third geometric stiffness
$[\mathbf{H}^{\text{g--3}}]$
collects the relations in Eqs.~(\ref{eq:equil1}),
(\ref{eq:db}), and~(\ref{eq:dw})
for all $N$ particles,
\begin{equation} \label{eq:Hg3}
\left.\begin{array}{ll}
\Dtp\times\Bp &= -\Dtp\times\Sum\Fpq \\
\Dtp\times\Wp &= -\Dtp\times\Sum\left(\Rpq\times\Fpq + \Mpq \right)
\end{array}
\right\}
\ \rightsquigarrow\ %
\LB \mathbf{H}^{\text{g--3}} \RB_{6N\times 6N}
\LB d\mathbf{u}\\: d\boldsymbol{\theta} \RB_{6N \times 1}\;.
\end{equation}
\subsection{Combined assembly stiffness matrix}\label{sec:combined}
Equation~(\ref{eq:A4}) can now be substituted into Eq.~(\ref{eq:prefinal})
to arrive at the stiffness relation for an assembly of
$N$ particles in the intended, target form of Eq.~(\ref{eq:H}):
\begin{equation}\tag{\ref{eq:H}}
\LB \mathbf{H} \RB_{6N \times 6N}
\DMut_{6N \times 1} = \DMbw_{6N \times 1}
\end{equation}
with
\begin{align}\label{eq:final1}
\LB \mathbf{H} \RB
&=
\left(\LB \Hgone \RB
+ \LB \Hgtwo \RB
+ \LB \Hgthree \RB\right)
+ \LB \Hm \RB \\
\label{eq:final2}
&= \LB \mathbf{H}^{\text{g}} \RB + \LB \Hm \RB \;.
\end{align}
The geometric stiffness $[\Hg]$ is combined from three parts,
which merely correspond to three steps in deriving $[\Hg]$.
\par
Each of the
$6N\times 6N$ stiffnesses 
in Eq.~(\ref{eq:final1}) can be constructed from
the $M$ corresponding $12\times 12$ contact stiffnesses.
That is, the $12\times 12$ stiffness in Eq.~(\ref{eq:piece})
for a single contact is the sum
of four $12\times 12$ contributions
that correspond to the matrices 
$[\Hgone ]$, 
$[\mathbf{H}^{\text{g--2}}]$, 
$[\mathbf{H}^{\text{g--3}}]$, and
$[ \Hm ]$
in Eq.~(\ref{eq:final1}).
We note, however, that the two submatrices
$[\mathbf{H}^{q\text{--}p}]$ and
$[\mathbf{H}^{q\text{--}q}]$ in Eq.~(\ref{eq:piece}) are formed
from the vectors $\mathbf{f}^{qp}$, $\mathbf{r}^{qp}$, and $\mathbf{n}^{qp}$,
etc. instead of their ``$pq$'' counterparts.
We also note that the inner product of $[\Hgone ]$,
$[\mathbf{H}^{\text{g--2}}]$, or $[ \Hm ]$
with any rigid-body motion
$\Dut^{\text{rigid}}$
will be zero, since these three
stiffnesses are constructed from the
contact deformations $\Dudef$ and $\Dtdef$,
which are zero for any rigid-body motion.
The product 
$[\mathbf{H}^{\text{g--3}}] \Dut^{\text{rigid}}$
might, however, not equal zero, an anomaly that is resolved
in Section~\ref{sec:rigid_rotation}.
\par
The assembly stiffness $[\mathbf{H}]$ in Eq.~(\ref{eq:final1})
embodies four stiffness components:
two geometric components, $[\Hgone ]$
and $[\mathbf{H}^{\text{g--2}}]$,
that depend upon the particle shapes (surface curvatures)
and upon the current contact forces;
a third geometric component $[\mathbf{H}^{\text{g--3}}]$ that
depends upon the particle size (the radial vectors $\Rpq$)
as well as upon the current contact forces;
and a mechanical component 
$[ \Hm ]$
that depends upon the contact stiffnesses.
The geometric stiffness $[\Hg]$
would be required to distinguish the different
incremental responses of the three clusters in
Fig.~\ref{fig:Geometry}.
Having derived the incremental stiffness $[\mathbf{H}]$,
we now consider two related matters that must
be resolved before applying $[\mathbf{H}]$ to questions
of stability, bifurcations, and softening.
\subsection{Cluster rotations}\label{sec:rigid_rotation}
Questions of stability and softening, discussed
in Section~\ref{sec:stability}, will depend upon 
second-order work quantities, specifically, on the
signs of inner products such as
\begin{equation} \label{eq:InnerProducts}
\DMbw^{\text{T}} \DMut
\quad \text{and} \quad
\LB \D\mathbf{b}\\: \D\mathbf{w} \RB^{\text{T}}
\LB \D\mathbf{u}\\: \D\boldsymbol{\theta} \RB \;,
\end{equation}
where $[\D\mathbf{b} / \D\mathbf{w}]$ and 
$[\D\mathbf{u} / \D\boldsymbol{\theta}]$ are defined in this section.
Although the first product (\ref{eq:InnerProducts}$_{1}$) is a standard matter 
in structural stability analysis,
structures and machines are usually attached to foundations or
chassis, so that rigid body motions are not explicitly considered.
To investigate the internal stability of a granular material, 
as manifested in a granular cluster or a representative volume element,
we must reconcile possible rigid modes of rotation,
particularly when the cluster is analyzed as being
independent of the surrounding material.
We refer to such granular sub-systems as ``isolated clusters,''
and the second product, Eq.~(\ref{eq:InnerProducts}$_2$),
is more appropriate for their analysis.
\par
\begin{figure}
\centering
\includegraphics[scale=0.90]{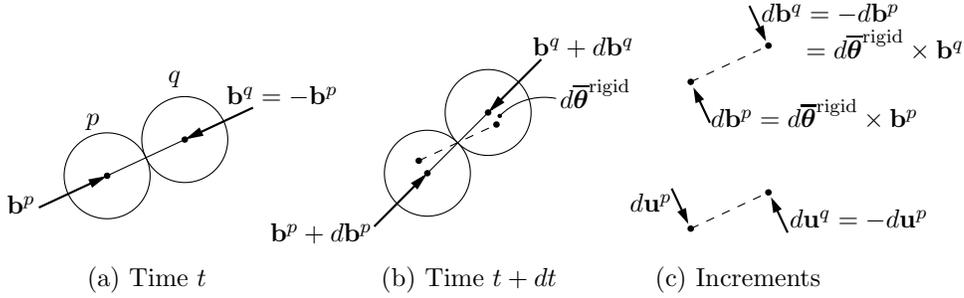}
\caption{Rigid rotation of an equilibrated system.}
\label{fig:rotate}
\end{figure}

Consider the isolated two-particle cluster in Fig.~\ref{fig:rotate}.
The particles are initially in equilibrium with the opposing external
forces $\mathbf{b}$ and $-\mathbf{b}$ (Fig.~\ref{fig:rotate}a).
The pair is then rotated in a rigid manner, along with its forces, 
through the angular increment $d\overline{\boldsymbol{\theta}}^{\text{rigid}}$,
as in Fig.~\ref{fig:rotate}b
(or, alternatively, the observer rotates by the angle
$-d\overline{\boldsymbol{\theta}}^{\text{rigid}}$).
The two increments of force, $d\mathbf{b}$ and $-d\mathbf{b}$,
are due entirely to the products
$d\overline{\boldsymbol{\theta}}^{\text{rigid}}\times \mathbf{b}$
and $-d\overline{\boldsymbol{\theta}}^{\text{rigid}}\times \mathbf{b}$
of Eq.~(\ref{eq:db}),
which are generated by the stiffness contribution
$[\mathbf{H}^{\text{g--3}}]$ of Eqs.~(\ref{eq:A4}) and~(\ref{eq:Hg3}).
The simpler inner product 
$[\mathbf{d\mathbf{b}}]^{\text{T}}[\mathbf{d\mathbf{u}}]$ equals $-2\,db\,du$
and is non-zero,
\emph{even though no second-order work involved}.
A stability criterion that is tied to this inner product
must obviously be amended to neglect such rigid rotation modes.
A similar situation arises in continuum theories of
internal instability and bifurcation, 
and these problems are typically corrected by using a corotational
or nominal stress rate in place of the Cauchy rate,
and by taking advantage of a symmetry of the stiffness tensor that negates
any spin component of the velocity gradient
\cite{Hill:1958a,Rice:1976a,Bazant:1991a}.
\par
When investigating the stability of a discrete system, 
certain corotational ``$\D$'' increments should be used,
as in the second inner product of Eq.~(\ref{eq:InnerProducts}).
To this end, 
we first derive a projection of the particle motions $\Dut$
onto the vector subspace of rigid rotations.
A rigid rotation of the entire system by an angle 
$d\overline{\boldsymbol{\theta}}^{\text{rigid}}$
produces the following motions, 
$d\mathbf{u}^{p,\overline{\boldsymbol{\theta}}}$ and
$d\boldsymbol{\theta}^{p,\overline{\boldsymbol{\theta}}}$,
of a single particle $p$ having the position $\mathbf{x}^{p}$
(Fig.~\ref{fig:particles}):
\begin{align}\label{eq:dutrigid}
d\mathbf{u}^{p,\overline{\boldsymbol{\theta}}} &=
  d\overline{\boldsymbol{\theta}}^{\text{rigid}} \times
  \mathbf{x}^{p} \\
d\boldsymbol{\theta}^{p,\overline{\boldsymbol{\theta}}} &=
  d\overline{\boldsymbol{\theta}}^{\text{rigid}} \;,
\end{align}
and the motions of all $N$ particles can be collected in a matrix
form as
\begin{equation}\label{eq:dtrigid}
\LB d\mathbf{u}^{\overline{\boldsymbol{\theta}}}\\:
    d\boldsymbol{\theta}^{\overline{\boldsymbol{\theta}}} \RB_{6N\times 1} =
\LB \mathbf{C} \RB_{6N\times 3}
\LB d\overline{\boldsymbol{\theta}}^{\text{rigid}} \RB_{3\times 1} \;.
\end{equation}
Conversely, the rigid rotation $d\overline{\boldsymbol{\theta}}^{\text{rigid}}$
of a system of $N$ moving particles can be extracted from their
$6N$ motions $\Dut$
by multiplying by
the Moore--Penrose inverse $[\mathbf{C}]^{+}$:
\begin{equation}\label{eq:Moore1}
\LB d\overline{\boldsymbol{\theta}}^{\text{rigid}} \RB_{3\times 1} =
\LB \mathbf{C} \RB_{3\times 6N}^{+}
\DMut_{6N\times 1}  \;,
\end{equation}
with
\begin{equation}
\LB \mathbf{C} \RB^{+} =
\left( \LB \mathbf{C} \RB^{\text{T}} \LB \mathbf{C} \RB\right)^{-1}
\LB \mathbf{C} \RB^{\text{T}} \;.
\end{equation}
The rigid-rotation mode 
$d\overline{\boldsymbol{\theta}}^{\text{rigid}}$
can then be removed
from the original particle motions $\Dut$
by projecting them onto the subspace that excludes
rigid rotations:
\begin{equation} \label{eq:ddu}
\LB \D\mathbf{u}\\: \D\boldsymbol{\theta} \RB_{6N\times 1} =
\LB \mathbf{P}^{\text{n--r--r}} \RB_{6N\times 6N}
\DMut_{6N\times 1} \;.
\end{equation}
The ``$\D$'' projected motions
$[\D\mathbf{u} / \D\boldsymbol{\theta}]$
are objective and contain no systematic rigid rotation
of the $N$ particles.
The ``no--rigid--rotation'' (n--r--r) projection matrix,
$[\mathbf{P}^{\text{n--r--r}}]$, is given by
\begin{equation}\label{eq:Pnrr}
\LB \mathbf{P}^{\text{n--r--r}} \RB_{6N\times 6N} =
\LB \mathbf{I} \RB_{6N\times 6N}
- \LB \mathbf{P}^{\text{r--r}} \RB_{6N\times 6N} \;,
\end{equation}
where the projection matrix $[\mathbf{P}^{\text{r--r}}]$
for ``rigid--rotations'' (r--r) is
\begin{equation}\label{eq:Prr}
\LB \mathbf{P}^{\text{r--r}} \RB =
\LB \mathbf{C} \RB \LB \mathbf{C} \RB^{+} \;.
\end{equation}
Both $[\mathbf{P}^{\text{r--r}}]$ and
$[\mathbf{P}^{\text{n--r--r}}]$ are symmetric and idempotent.
\par
The stiffness relation in Eq.~(\ref{eq:H}) can be rewritten by
substituting the motions
$[\D\mathbf{u} / \D\boldsymbol{\theta}]$
in Eq.~(\ref{eq:ddu})
for the motions
$[d\mathbf{u} / d\boldsymbol{\theta}]$:
\begin{equation}\label{eq:equilibrium5}
\LB \mathbf{H} \RB
\LB \mathbf{P}^{\text{n--r--r}} \RB
\DMut
= \LB d\mathbf{b}\\: d\mathbf{w} \RB
- \LB \mathbf{H} \RB \LB \mathbf{P}^{\text{r--r}} \RB
  \DMut \;.
\end{equation}
The proper use of $[\mathbf{C}]$ and its related matrices requires that
the particle positions $\Xp$ in Eq.~(\ref{eq:dutrigid})
are measured from the center of the $N$-particle cluster,
so that $\sum_{i=1}^{N}\Xp=0$.
By choosing another origin, 
the product $[\mathbf{P}^{\text{r--r}}]\Dut$ will improperly
deal with rigid body translations, producing an apparent (and false) rotation
of the system.
If another origin must be used, three additional columns
should be appended to the matrix $[\mathbf{C}]$, so that
the column space of $[\mathbf{C}]$ spans both rigid rotations
$d\overline{\boldsymbol{\theta}}^{\text{rigid}}$ and rigid translations
$d\overline{\mathbf{u}}^{\text{rigid}}$.
The following derivations use a central origin and
the simpler $6N\times 3$
matrix $[\mathbf{C}]$ of Eq.~(\ref{eq:dtrigid}).
\par
Equation~(\ref{eq:equilibrium5}) is an alternative
to Eq.~(\ref{eq:H}),
and it effects two changes that are relevant
to stability analysis.
First, the product
$[\mathbf{P}^{\text{n--r--r}}][d\mathbf{u} / d\boldsymbol{\theta}]=[\D\mathbf{u} / \D\boldsymbol{\theta}]$
on the left of Eq.~(\ref{eq:equilibrium5})
removes rigid modes of rotation from the full $\mathbb{R}^{6N}$
space of particle motions $\Dut$.
As such, the non-zero movements $d\mathbf{u}$ and $-d\mathbf{u}$
in Fig.~\ref{fig:rotate}c would be replaced with 
$\D\mathbf{u}=-\D\mathbf{u}=0$.
Second, the force increments
$[d\mathbf{b}/d\mathbf{w}]$ on the right of Eq.~(\ref{eq:equilibrium5})
are reduced by the increments that are produced merely by a systematic
rigid rotation of the $N$ particles.
The matrix $[\mathbf{H}]$ in Eq.~(\ref{eq:equilibrium5}) 
is the sum of the four contributions given in
Eq.~(\ref{eq:final1}), but three of these contributions originate solely from
the objective contact deformations $\Dudef$ and $\Dtdef$:
the matrices $[\Hgone ]$,
$[ \Hgtwo ]$, and 
$[ \Hm ]$, as defined in
Eqs.~(\ref{eq:Hg1}), (\ref{eq:A3}), and~(\ref{eq:constitMatrix}).
These three contributions are unaffected by a systematic rigid rotation
of the assembly.
For example, 
with the ``$\text{g--1}$'' contribution,
the product on the right of Eq.~(\ref{eq:equilibrium5}) is
$[\Hgone ][\mathbf{P}^{\text{r--r}}] \Dut =0$.
Only the $[\mathbf{H}^{\text{g--3}}]$ contribution
is affected by a rigid rotation,
as is seen by substituting a systematic rotation
$d\overline{\boldsymbol{\theta}}^{\text{rigid}}$
into the definition in Eq.~(\ref{eq:Hg3}).
\par
We define the
force increments $\D\mathbf{b}$ and $\D\mathbf{w}$
as the expression on the right of Eq.~(\ref{eq:equilibrium5}),
which can also be written in the alternative forms
\begin{equation} \label{eq:ddb}
\begin{split}
\DDMbw &\ =\ %
\DMbw
- \LB \mathbf{H} \RB \LB \mathbf{P}^{\text{r--r}} \RB
  \DMut
\ =\ %
\DMbw - \LB \Hgthree \RB
\LB \mathbf{P}^{\text{r--r}} \RB
\DMut
\\
&\ =\ %
\DMbw
- \LB \Hgthree \RB \LB \mathbf{C} \RB
\LB d\overline{\boldsymbol{\theta}}^{\text{rigid}} \RB \;.
\end{split}
\end{equation}
That is,
the force increments $[d\mathbf{b}/d\mathbf{w}]$
are reduced by a common rotation of the current 
external forces $\Bp$ and $\Wp$,
a rotation that produces the increments
$d\overline{\boldsymbol{\theta}}^{\text{rigid}}\times\Bp$ and
$d\overline{\boldsymbol{\theta}}^{\text{rigid}}\times\Wp$ for $p=1\ldots N$.
The forces $d\Bp$ and $d\Bq$ in Fig.~\ref{fig:rotate}
would be eliminated by the subtracted 
terms in Eq.~(\ref{eq:ddb}).
Increments $\D\mathbf{b}$ and $\D\mathbf{w}$ are objective.
\par
We define the modified stiffnesses $[\mathbbm{H}]$ and
$[\HHHH]$ as
\begin{equation}\label{eq:newstiffnesses}
\LB \mathbbm{H} \RB =
\LB \mathbf{H} \RB \LB \mathbf{P}^{\text{n--r--r}} \RB\;,\quad
\LB \HHHH \RB =
\LB \mathbf{P}^{\text{n--r--r}}\RB^{\text{T}}
    \LB\mathbf{H} \RB \LB \mathbf{P}^{\text{n--r--r}} \RB\;.
\end{equation}
When combined with
the definitions~(\ref{eq:equilibrium5}) and~(\ref{eq:ddb}),
the stiffness relation~(\ref{eq:H}) can be written in the following
alternative forms
\begin{equation} \label{eq:problem}
\LB \mathbf{H} \RB \DMut
  = \DMbw \quad\text{or}\quad
\LB \mathbbm{H} \RB \DMut
  = \DDMbw \;.
\end{equation}
Possible bifurcations in an isolated granular cluster
are resolved by seeking multiple solutions of the second form 
(Section~\ref{sec:uniqueness}).
The possible instability or softening of an isolated granular
cluster is resolved by considering the following inner
product:
\begin{equation}\label{eq:newinner}
\DDMbw^{\text{T}}
\LB \D\mathbf{u}\\: \D\boldsymbol{\theta} \RB
=
\DMut^{\text{T}}
\LB \HHHH \RB
\DMut \;,
\end{equation}
as discussed in Section~\ref{sec:softening}.
Because the projection matrix $[\mathbf{P}^{\text{n--r--r}}]$
is symmetric and idempotent, the two matrices
$[\HHH]$ and $[\HHHH]$ share the same eigenvalues.
This characteristic is proven by supposing 
that $\lambda$ and $[\boldsymbol{\nu}]$ are an eigenvalue
and eigenvector of $[\HHH]$:
\begin{align} \label{eq:proof1}
\LB \HHH \RB \LB \boldsymbol{\nu} \RB &= \lambda\LB \boldsymbol{\nu} \RB \\
\LB \HHHH \RB \LB \boldsymbol{\nu} \RB &=
  \lambda\LB\mathbf{P}^{\text{n--r--r}}\RB\LB \boldsymbol{\nu} \RB\\
\label{eq:proof3}
\LB \HHHH \RB\LB\mathbf{P}^{\text{n--r--r}}\RB\LB \boldsymbol{\nu} \RB
&=
\lambda\LB\mathbf{P}^{\text{n--r--r}}\RB\LB \boldsymbol{\nu} \RB
\end{align}
where we have substituted Eq.~(\ref{eq:newstiffnesses}) between
the first and second expressions and have used the idempotent property
of $[\mathbf{P}^{\text{n--r--r}}]$ to arrive at the third expression.
The result shows that $\lambda$ is also an eigenvalue of
$[\HHHH]$, but with the associated eigenvector
$[\mathbf{P}^{\text{n--r--r}}][\boldsymbol{\nu}]$.
Stability depends, however, upon the eigenvalues of the symmetric part
of $[\HHHH]$, which might differ from those of
$[\HHHH]$ itself or of the symmetric part of
$[\HHH]$ (Section~\ref{sec:softening}).
\subsection{Elastic-plastic contact stiffness}\label{sec:epstiffness}
\citeN{Michalowski:1978a} and \shortciteN{Radi:1999a} have derived a
simple contact stiffness by applying concepts of elasto-plasticity
theory.
We briefly review this stiffness, as it will serve as a prototype
for investigating the stability and softening of particle
sub-regions (Section~\ref{sec:stability}).
The contact stiffness is incrementally nonlinear 
with two branches: 
an elastic branch that is characterized
with the normal and tangential stiffnesses
$k^{pq}$ and $\alpha k^{pq}$,
and a sliding branch characterized by a friction coefficient
$\mu^{pq}$.
Whenever sliding becomes possible,
the active branch is
determined by the direction of the contact
deformation $\Dudef$.
Sliding occurs at a firm contact when two conditions are met:
\begin{enumerate}
\item
When the current contact force satisfies the 
yield condition $Q^{pq}=0$: 
\begin{equation}\label{eq:yield}
Q^{pq} = Q(\Fpq ) =
\left |
\Fpq - (\Npq \cdot \Fpq) \Npq
\right |
+
\mu \Fpq \cdot \Npq
= 0 \;.
\end{equation}
This yield condition depends upon the current contact force $\Fpq$,
which is known \emph{a priori}.
With the isotropic frictional behavior
in Eq.~(\ref{eq:yield}), the yield condition is axisymmetric
within the contact plane
(see \citeNP{Michalowski:1978a} for alternative, asymmetric forms).
\item
When the contact deformation $\Dudef$ is directed outward
from the yield surface in displacement space, the condition
$S^{pq}>0$:
\begin{equation}\label{eq:flow}
S^{pq} = S(\Fpq , \Dudef) = \Gpq \cdot \Dudef > 0 \;,
\end{equation}
where the yield surface $Q$ has the normal direction
\begin{equation}\label{eq:G}
\Gpq = k \left( \alpha \Hpq + \mu \Npq \right)
\end{equation}
and the unit sliding direction $\Hpq$ is tangent to the contact plane
and aligned with the current contact force $\Fpq$:
\begin{equation}\label{eq:Hdirection}
\Hpq = \frac{  \Fpq - (\Npq \cdot\Fpq ) \cdot \Npq   }
            {| \Fpq - (\Npq \cdot\Fpq ) \cdot \Npq |} \;.
\end{equation}
\end{enumerate}
With this simple model and a hardening modulus of zero, 
the contact stiffness tensor $\FFpq$
in Eq.~(\ref{eq:constitutive}) has two branches,
elastic and sliding, given by
\begin{equation}\label{eq:ContactF}
\FFpq = \begin{cases}
\mathbf{F}^{pq\text{, elastic}} =
  k\left[ \alpha\mathbf{I} + (1-\alpha)\Npq\otimes\Npq \right]
  & \text{if } Q^{pq}<0 \text{ or } S^{pq}\leq 0 \\
  \mathbf{F}^{pq\text{, sliding}} =
  \mathbf{F}^{pq\text{, elastic}} -
  \Hpq\otimes\Gpq
  & \text{if } Q^{pq}=0 \text{ and } S^{pq}>0
\end{cases}
\end{equation}
where $\mathbf{I}$ is the Kronecker, identity tensor.
Because the sliding and yield directions do not coincide
($\Hpq \neq \Gpq$), sliding is non-associative and
the contact stiffness in Eq.~(\ref{eq:ContactF}$_2$) is
asymmetric and may lead to negative second-order work at the contact.
The sliding behavior possesses deviatoric associativity, however,
since the sliding direction $\Hpq$ is aligned with
the tangential component of the yield surface normal
$\Gpq$ \cite{Bigoni:2000a}.
The yield condition in Eq.~(\ref{eq:yield}) will likely
be met at multiple contacts within a granular assembly,
which will lead to a combined stiffness $\mathbf{H}^{m}(\Dut)$ that is
incrementally nonlinear and has multiple stiffness branches
(Section~\ref{sec:stability}).
\par
The derivation of Eq.~(\ref{eq:ContactF}) assumes that the
two particles are in firm contact, as opposed to grazing contact
\shortcite{Radi:1999a}.
For a firm contact, the incremental stiffness
is piece-wise linear, having linear behavior within each 
branch of Eq.~(\ref{eq:ContactF}).
Grazing contacts have thoroughly nonlinear behavior and are not treated
further in this work.
\section{Uniqueness, internal stability, and softening}\label{sec:stability}
With a typical structural system, questions of
uniqueness and stability can be resolved by investigating the
determinant and eigenvalues of its stiffness matrix.
Although we can use this approach with granular systems,
the incremental analysis will likely be complicated by two conditions:
(1) incrementally nonlinear stiffnesses $\HHH$ 
and $\HHHH$ having multiple branches,
and (2) the asymmetry of these stiffnesses.
Both factors are now considered.
We confine this study, however, to isolated particle clusters,
which lack any displacement constraints that would otherwise
prevent rigid motions of the cluster, and
the more general problem of constrained granular
systems is left for future study.
With isolated clusters, the matrices $[\HHH]$ and $[\HHHH]$ in
Eqs.~(\ref{eq:problem}$_2$) and~(\ref{eq:InnerProducts}$_2$)
will be examined in place of matrix 
$[\mathbf{H}]$ and Eqs.~(\ref{eq:H}) and~(\ref{eq:InnerProducts}$_1$),
and the inevitable (but less interesting) 
rigid-body motions will be referred to as
\emph{trivial solutions} of Eq.~(\ref{eq:problem}$_2$).
\par
The geometric stiffness $[\Hg]$ of smooth particles is independent of the
loading direction,
but the mechanical stiffness $\Hm(\Dut )$ can be incrementally
nonlinear, having a finite number $L$ of stiffness branches,
represented by the matrices
$[\mathbf{H}^{\text{m, } 1}]$,
$[\mathbf{H}^{\text{m, } 2}]$,
$[\mathbf{H}^{\text{m, } 3}]$, $\ldots\,$, %
$[\mathbf{H}^{\text{m, } L}]$.
Because the contact behavior is assumed homogeneous of degree one
(Eqs. \ref{eq:constitutive}--\ref{eq:constitutiveM}),
the active branch of $\Hm(\Dut)$ is determined by the unit loading
direction $\Dut/\left| \Dut \right|$.
Although incrementally nonlinear, we assume that the 
incremental mapping 
$\Hm:\,\Dut\rightarrow[d\mathbf{b}/ d\mathbf{w}]$
is continuous and piece-wise linear, so that two adjacent branches share the
same stiffness along their shared boundary, and the behavior is linear within 
each branch.
The example contact model in Section~\ref{sec:epstiffness}
would lead to incrementally nonlinear mappings
$\Hm(\Dut)$ having these characteristics.
With this contact model,
a single contact has one stiffness if it is elastic
($Q<0$ in Eq.~\ref{eq:yield}), but it has two branches
when the yield surface has been reached.
If $M^{\text{s}}$ of the $M$ contacts are known to
be potentially sliding, having a current $Q=0$, then the combined stiffness
$\Hm(\Dut )$ has $L=2^{M^{\text{s}}}$ branches.
The active branch is determined by applying $M^{\text{s}}$ 
independent sliding conditions,
each in the form of Eq.~(\ref{eq:flow}).
\par
The $i$th stiffness branches $[\HH^{i}]$, $[\HHH^{i}]$, and $[\HHHH^{i}]$ 
will often be asymmetric.
Symmetry of the mechanical stiffness $[\Hm ]$ depends upon the symmetry
of the individual contact stiffnesses---the $\FFpq$ and $\MMpq$
in Eqs.~(\ref{eq:constitutive}) and~(\ref{eq:constitutiveM})---whose 
symmetry is lost when contacts begin to slide.
The geometric stiffness $[\Hg ]$ is symmetric only if all $M$ contact forces
lack a tangential component.
\subsection{Uniqueness}\label{sec:uniqueness}
We now consider whether Eq.~(\ref{eq:problem}$_2$)
admits multiple non-trivial solutions for a given
force increment $[\D \mathbf{b}/\D \mathbf{w}]$.
For a \emph{linear} and possibly asymmetric structural system that is
constrained from rigid-body motions,
uniqueness is assured when the determinant 
$\text{det}([\HH ]) \neq 0$ or, alternatively, when
$[\HH ]$ has no eigenvalues that are zero.
Isolated granular clusters are linear when no contacts are yet sliding,
but even then,
the usual criterion must be modified to exclude rigid
motions of the cluster as possible bifurcation modes. 
Using the stiffness $[\HHH]$
of Eq.~(\ref{eq:newstiffnesses}) in place of $[\HH]$,
an isolated \emph{linear} granular cluster admits no non-trivial
bifurcations when $[\HHH ]$ has only six eigenvalues that
are zero---the eigenvalues that correspond to the six independent
rigid-body motions.
A seventh zero-eigenvalue signals a 
condition of \emph{neutral equilibrium} and the presence
of non-trivial, bifurcating solutions of the linear
equations. 
In this case, 
any multiple of the seventh eigenvector $[\boldsymbol{\nu}^{(7)}]$
can be added to a solution of the non-homogeneous
Eq.~(\ref{eq:problem}$_2$) to produce a family of solutions.
\par
When contacts are sliding, granular behavior is inelastic 
and incrementally nonlinear, and multiple
branches of the stiffness $\HHH(\Dut)$ must be considered for admitting 
solutions of Eq.~(\ref{eq:problem}$_2$).
For an isolated cluster, non-uniqueness arises when two non-trivial
solutions, $\Dut ^{a}$ and $\Dut ^{b}$, exist:
\begin{equation} \label{eq:nonunique}
\LB \HHH^{a}\RB \DMut^{a} = \DDMbw
\quad\text{and}\quad
\LB \HHH^{b}\RB \DMut^{b} = \DDMbw
\end{equation}
where the difference $\Dut ^{a} - \Dut ^{b}$ is not a rigid-body motion,
and where the two stiffness branches
$[\HHH^{a}]$ and $[\HHH^{b}]$ are consistent with the directions
of their solution vectors $\Dut^{a}$ and $\Dut^{b}$, respectively.
By \emph{consistent} we mean that a product
$[\HHH^{i}]\Dut$ involves motions $\Dut$ 
that lie within the particular domain of the
branch $[\HHH^{i}]$, which could be verified by checking $M^{\text{s}}$
sliding conditions in the form of Eq.~(\ref{eq:flow}).
The non-uniqueness in Eq.~(\ref{eq:nonunique}) can arise in two
ways:
\begin{enumerate}
\item
\emph{Type~1 non-uniqueness} occurs
when $\Dut^{a}$ and $\Dut^{b}$ belong to different branches of
the stiffness $\HHH(\Dut)$, such that $[\HHH^{a}]\neq[\HHH^{b}]$.
\item
\emph{Type~2 non-uniqueness} occurs
when a single branch, say $[\HHH^{a}]$
with solution $\Dut^{a}$, satisfies 
Eq.~(\ref{eq:nonunique}$_{1}$) and has a seventh eigenvalue that is
zero.
Because behavior within each branch is assumed to be linear,
a family of non-trivial solutions
$\Dut ^{b} = \Dut ^{a} + \gamma[\boldsymbol{\nu}^{(7)}]$
is associated with the solution 
$\Dut ^{a}$ (although the scalar $\gamma$ may need to
be restricted to keep $\Dut ^{b}$ within the same branch
as $\Dut ^{a}$).
\end{enumerate}
The first situation is possible when some of the contact stiffnesses
$\FFpq$ are not positive definite, as with the sliding contacts
of Eq.~(\ref{eq:ContactF}$_{2}$).
In this case, the Hill condition
$(\Dut ^{a} - \Dut ^{b})^{\text{T}}
([\HHH^{a}]\Dut^{a} - [\HHH^{b}]\Dut^{b})>0$
might not be met for certain vectors
$\Dut ^{a}$ and $\Dut ^{b}$,
which can permit Type~1 non-uniqueness.
\par
The two types of non-uniqueness suggest an algorithm for seeking
possible bifurcating solutions of Eq.~(\ref{eq:problem}$_{2}$).
For the given loading $[\D\mathbf{b}/\D\mathbf{w}]$,
each of the $L=2^{M^{\text{s}}}$ branches of $[\HHH^{i}]$, $i=1\ldots L$, 
must be checked for a possible solution to Eq.~(\ref{eq:problem}$_{2}$).
If a solution appears to exist within the particular branch
$[\HHH^{i}]$, this
solution $\Dut$ must also be checked for its consistency with
the loading conditions of that branch 
(e.g., by applying Eq.~\ref{eq:flow} to each of the $M^{s}$
potentially sliding contacts).
If multiple branches give non-trivial and consistent solutions,
then Type~1 non-uniqueness is present.
The number of zero-eigenvalues must also be counted for
each branch that yields a non-trivial and consistent solution.
If the matrix of any solution branch 
has more than six zero-eigenvalues with consistent
eigenvectors, then Type~2
non-uniqueness is present. 
\subsection{Stability and softening}\label{sec:softening}
We adopt the usual criterion of stability for time-invariant
systems:
a system is stable if positive work is required for all load increments
that maintain equilibrium
\cite{Kratzig:1995a,Petryk:2000a}.
If an isolated granular cluster is already in equilibrium under
the current external forces $[\mathbf{b}/\mathbf{w}]$,
then \emph{the system is stable} if the second-order work is positive
for all increments $\Dut$:
\begin{equation} \label{eq:stab}
\left(\DMut^{\text{T}}\LB \HHHH^{i} \RB \DMut > 0,\; 
\forall \DMut \text{ consistent with } 
\LB\HHHH^{i}\RB\right),\;i=1\ldots L
\;\Rightarrow
\text{Stability}
\end{equation}
where the inner product in Eqs.~(\ref{eq:InnerProducts}$_{2}$) 
and~(\ref{eq:newinner})
is used in place of Eq.~(\ref{eq:InnerProducts}$_{1}$).
In verifying condition~(\ref{eq:stab}),
all branches $i=1\ldots L$ must be checked, and with each
branch, all consistent vectors $\Dut$ must be checked.
The loading direction $\Dut$
must be consistent with the particular branch $[\HHHH^{i}]$
that is being checked.
The condition (\ref{eq:stab}), however, 
is sufficient but not necessary for stability,
since higher-order work terms are not considered in this study.
In the stability of Eq.~(\ref{eq:stab}), 
a stable cluster can sustain the current dead
load $[\mathbf{b}/\mathbf{w}]$, 
insofar as small disturbances $[\D\mathbf{b}/\D\mathbf{w}]$
produce only small displacements.
\par
Conditions for \emph{neutral stability} and \emph{instability} are likewise
given by the criteria
\begin{align}
\label{eq:neutral}
&\text{Neutral stability}\;\Rightarrow
\exists\;\text{n.t.}\DMut
\text{ consistent with }\LB\HHHH^{i}\RB, \; 
\DMut^{\text{T}}
\LB \HHHH^{i} \RB \DMut = 0 \\
\label{eq:instab}
&\exists\DMut
\text{ consistent with }\LB\HHHH^{i}\RB, \;
\DMut^{\text{T}}\LB \HHHH^{i} \RB \DMut < 0
\;\Rightarrow \text{Instability}
\end{align}
(e.g., \citeNP{Bazant:1991a}),
where ``n.t.'' denotes a non-trivial displacement---one that does
not lie in the sub-space of rigid-body motions
(Section~\ref{sec:rigid_rotation}).
As with Eq.~(\ref{eq:stab}),
$[\HHHH^{i}]$ must be consistent with the displacement $\Dut$
that is being tested.
Once unstable, a granular system becomes dynamic and the particles'
inertias influence their subsequent motions, unless, of course,
some of the motions in $\Dut$ are externally constrained.
\par
\emph{Softening} occurs in any loading direction
$\Dut$, perhaps constrained, that produces negative second-order work,
as in Eq.~(\ref{eq:instab}) (e.g., \citeNP{Valanis:1985a}).
\par
The stability conditions in Eqs.~(\ref{eq:stab})--(\ref{eq:instab})
are determined, of course, by the symmetric
part $[\widehat{\HHHH}^{i}]$ of the stiffness
$[\HHHH^{i}]$,
where 
$[\widehat{\HHHH}^{i}] = (1/2)([\HHHH^{i}] + [\HHHH^{i}]^{\text{T}})$.
These stability conditions
differ from the uniqueness criterion
in Section~\ref{sec:uniqueness}, since the
latter depends upon the determinant or eigenvalues of the full,
asymmetric stiffness $[\HHH^{i}]$ 
(or of $[\HHHH^{i}]$, since $[\HHH^{i}]$ and $[\HHHH^{i}]$
share the same eigenvalues, Eqs.~\ref{eq:proof1}--\ref{eq:proof3}).
Because the smallest real eigenvalue of $[\widehat{\HHHH}^{i}]$ is no greater
than the smallest real eigenvalue of $[\HHHH^{i}]$,
instability does not imply a loss of uniqueness.
On the other hand, the neutral equilibrium of Type~2 non-uniqueness
implies neutral stability, since
$[\HHH]\Dut=0\;\Rightarrow\;\Dut^{\text{T}}[\HHHH]\Dut=0$.
That is, a granular cluster can be unstable and soften before
passing through neutral equilibrium.
\par
The definitions in Eqs.~(\ref{eq:stab})--(\ref{eq:instab}) suggest
an algorithm for investigating the stability of an isolated
granular cluster.
Each of the $L=2^{M^{s}}$ branches $[\HHHH^{i}]$,
$i=1\ldots L$, are examined by
finding the eigenvalues of their symmetric parts $[\widehat{\HHHH}^{i}]$.
At least six eigenvalues will be zero for every
$[\widehat{\HHHH}^{i}]$, corresponding to its rigid-body modes.
A \emph{sufficient condition for stability} is that all
branches $[\widehat{\HHHH}^{i}]$ have only positive eigenvalues, 
except for the six zero-eigenvalues.
A \emph{sufficient condition for neutral stability or instability}
is the presence of a seventh zero-eigenvalue
or a negative eigenvalue, respectively,
provided that the corresponding eigenvector is consistent with the
presumed loading conditions of the branch
(i.e., by applying Eq.~\ref{eq:flow} to each 
of the $M^{s}$ potentially sliding contacts).
If the eigenvector is consistent, then it represents an eigenmode
of neutral stability or of instability, respectively.
\par
The sufficient conditions in this algorithm can be readily
applied by examining the eigenvalues and eigenvectors
of all branches $[\HHHH^{i}]$, $i=1\ldots L$.
Implementation details are provided in
Appendix~\ref{app:eigen}.
The algorithm, however, provides a criterion that is over-sufficient
(i.e. not necessary) for instability:
even though all consistent eigenvectors of
a branch $[\widehat{\HHHH}^{i}]$ may have positive eigenvalues,
a \emph{non-consistent} eigenvector having a negative eigenvalue
might be linearly combined with a consistent eigenvector 
to produce a consistent motion $\Dut$ that brings about
a negative inner product in Eq.~(\ref{eq:instab}).
Likewise, the algorithm provides conditions that are
over-sufficient for stability:
a negative eigenvalue might exist, 
but if its corresponding eigenvector is non-consistent, the
presence of the negative eigenvalue does not imply instability.
\section{Examples}\label{sec:example}
\subsection{Two-particle system}\label{sec:twoparticle}
\begin{figure}
\centering
\includegraphics[scale=0.70]{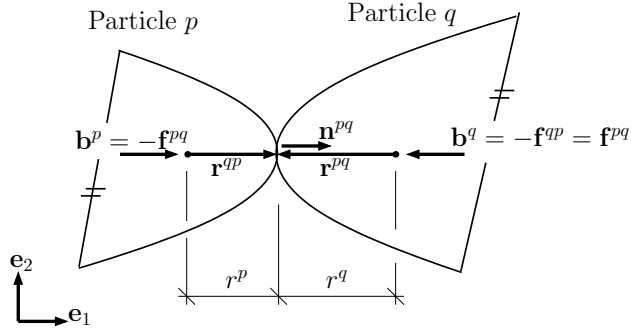}
\caption{An example two-particle cluster.}
\label{fig:Example}
\end{figure}

We consider an isolated cluster of two particles, ``$p$'' and ``$q$'',
and investigate its stability (Fig.~\ref{fig:Example}).
The example system is simplified with the following four restrictions:
\begin{enumerate}
\item
Motions are restricted to the $x_{1}$--$x_{2}$ plane, with the basis
vectors $\Eone$ and $\Etwo$.
\item
The radial vectors $\Rpq$ and $\Rqp$ are collinear, 
such that 
$\Xp$, $\Xq$, and the contact point lie on a common line.  
The radii $\Rpq$ and $\Rqp$ are oriented along the $\Eone$ direction.
\item
The contact normal $\Npq$ is aligned with the radii $\Rpq$ and $\Rqp$.
\item
No body moments are applied ($\Wp = \Wq = \Zero$), so that
the current body forces, $\Bp$ and $\Bq$, are collinear and
self-equilibrating: $\Bp = -\Bq$.
\end{enumerate}
We also adopt the simple contact model
of Section~\ref{sec:epstiffness}, and neglect any contact moment
resistance ($\mathfrak{d}\Mpq = -\mathfrak{d}\Mqp = \Zero$
in Eq.~\ref{eq:constitutiveM}).
Because the contact force $\Fpq$ is entirely normal,
the contact stiffness is elastic, as in Eq.~(\ref{eq:ContactF}$_{1}$):
\begin{equation}\label{eq:simple}
\DFfpq = 
  k \left[ \alpha\mathbf{I} + (1-\alpha)\Npq\otimes\Npq\right] \cdot \Dudef\;,
\end{equation}
where the positive
stiffnesses $k$ and $\alpha k$ are in the normal and tangential directions.
The particles are pressed together with a current 
compressive normal force $f$,
and the two particles have the convex radii of curvature 
$\Rhop$ and $\Rhoq$ at their contact.
\par
The stiffness $[\HH]$ for the two-particle system is derived
in Appendix~\ref{app:example} with the following result:
\begin{align}\label{eq:exampleH}
\LB\HH\RB &\DUa = \left( \LB\HH^{m} \RB + \LB\HH^{g} \RB \right) \DUa \\
\label{eq:exampleH2}
=
&\left(k
\left[ \begin{MAT}(r)[2pt]{ccc:ccc}
1 & 0 & 0 & -1 & 0 & 0 \\
0 & \alpha & \alpha r^p & 0 & -\alpha & \alpha r^q \\
0 & \alpha r^p & \alpha (r^p)^2 & 0 & -\alpha r^p & \alpha r^p r^q \\:
-1 & 0 & 0 & 1 & 0 & 0 \\
0 & -\alpha    & -\alpha r^p    & 0 &  \alpha     & -\alpha r^q \\
0 & \alpha r^q & \alpha r^p r^q & 0 & -\alpha r^q & \alpha (r^q)^2 \\
\end{MAT} \right] \right.
\\ 
&\left. +\frac{f}{\Rhopq}
\left[ \begin{MAT}(r)[3pt]{ccc:ccc}
0 & 0 & 0 & 0 & 0 & 0 \\
0 & -1 & \Rhop - r^p & 0 & 1 & \Rhoq - r^q \\
0 & \Rhop-r^p & (\Rhop-r^p)(\Rhoq+r^p) & 0 & r^p-\Rhop & 
                                                   (\Rhoq-r^q)(r^p - \Rhop)\\:
0 & 0 & 0 & 0 & 0 & 0 \\
0 & 1 & r^p - \Rhop & 0 & -1 & r^q - \Rhoq \\
0 & \Rhoq-r^q & (\Rhoq-r^q)(r^p - \Rhop) & 0 & r^q-\Rhoq &
                                                   (\Rhoq-r^q)(\Rhop+r^q)\\
\end{MAT} \right] \right)
\LB du^p_1\\du^p_2 \\ d\theta^p_3\\: du^q_1\\du^q_2 \\ d\theta^q_3 \RB
\notag
\end{align}
Rather than give the full $12\times 12$ stiffness matrix for the pair,
we have discarded the $\Ethree$ translation and the $\Eone$ and $\Etwo$
rotations and have derived the remaining $6 \times 6$ stiffness components.
The rows of matrix $[\mathbf{H}]$ are arranged
to produce forces $[d\mathbf{b}/d\mathbf{w}]$ in the following order:
$[db^p_1, db^p_2, dw^p_3, db^q_1, db^q_2, dw^q_3]^{\text{T}}$.
Both the mechanical and geometric stiffnesses are symmetric,
since
the mechanical stiffness is entirely elastic, and the
contact force lacks a tangential component.
The relative importance of the geometric and mechanical stiffnesses
is seen to depend upon the force-to-stiffness ratio $f/k$.
Moreover, if the two particles fit together like hand-in-glove,
with $\rho^{p}\approx -\rho^{q}$,
the quotient $f/(\rho^{p} + \rho^{q})$ is large, and the
geometric stiffness will dominate.
\par
\begin{figure}
\centering
\parbox[b]{3.4in}{\centering%
               \includegraphics[scale=0.90]{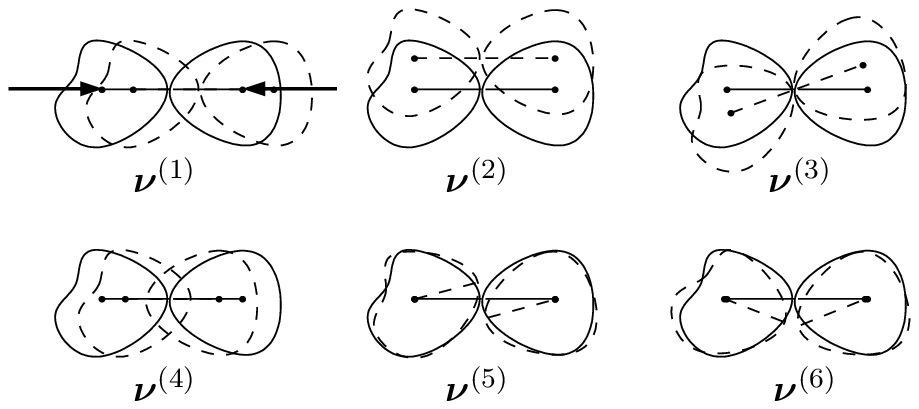}\\
               (a) Displacement modes}
\quad\quad\quad\quad
\parbox[b]{1.1in}{\centering%
\includegraphics[scale=0.90]{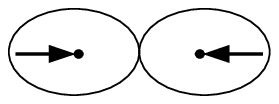}\\
(b) {\small Unstable} \\[0.20in]
\includegraphics[scale=0.90]{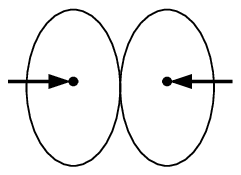}\\
(c) {\small Stable}}
\caption{Displacement modes and stability of two-particle systems.}  
\label{fig:Modes}
\end{figure}
\begin{figure}
\centering
\includegraphics[scale=0.90]{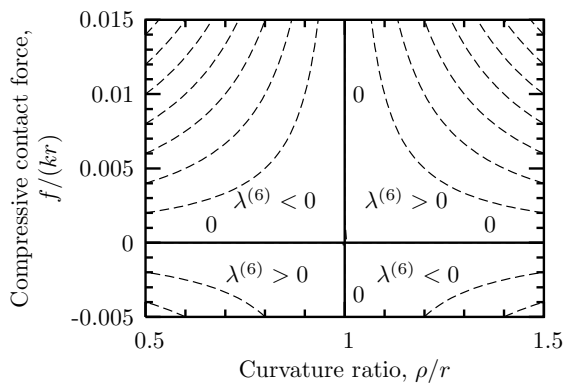}
\caption{Contour plot of the eigenvalue $\lambda^{(6)}$ when
$r^p=r^q$, $\rho^p=\rho^q$, and $\alpha=1$.}
\label{fig:Contours}
\end{figure}

Stability is investigated by finding the six eigenvalues
$\lambda^{(j)}$ of the matrix
$[\HHH]=[\mathbf{H}][\mathbf{P}^{\text{n--r--r}}]$,
where the projection $[\mathbf{P}^{\text{n--r--r}}]$ is computed from
the rotation vector 
$[\mathbf{C}]$ given
in Eq.~(\ref{eq:exampleC}) of Appendix~\ref{app:example}.
General expressions for some eigenvectors are too lengthy to
present here, but we make the following observations:
\begin{enumerate}
\item
Three eigenvalues are zero, corresponding to two
rigid translations and a rigid rotation (the eigenvectors
$\boldsymbol{\nu}^{(1)}$, $\boldsymbol{\nu}^{(2)}$,
and $\boldsymbol{\nu}^{(3)}$ in Fig.~\ref{fig:Modes}a).
\item
A fourth eigenvalue $\lambda^{(4)}$ is a positive $2k$, 
corresponding to the mode of normal contact indentation
($\boldsymbol{\nu}^{(4)}=[1/\sqrt{2}, 0, 0, -1/\sqrt{2}, 0, 0]^{\text{T}}$).
\item
Another positive eigenvalue corresponds to a tangential shearing at the 
contact (mode $\boldsymbol{\nu}^{(5)}$ in Fig.~\ref{fig:Modes}a).
\item
A sixth eigenvalue $\lambda^{(6)}$ can be positive, zero,
or negative depending on the radii and curvatures of the particles,
the two contact stiffnesses $k$ and $\alpha k$, and the force $f$.
\end{enumerate}
The sixth mode $\boldsymbol{\nu}^{(6)}$ is
the most interesting and corresponds to a rolling of the
particles at their contact (Fig.~\ref{fig:Modes}a).  
This mode can be investigated by restricting the two particles to
the same size and shape, with $r^{p}=r^{q}$ and $\rho^{p}=\rho^{q}$
at their contact.
Figure~\ref{fig:Contours} is a contour plot of
the sixth eigenvalue 
$\lambda^{(6)}$ for various combinations of curvature $\rho$
and compressive force $f$.
The dimensionless curvature $\rho/r$ ranges from
shapes that are relatively ``sharp'' 
($\rho/r < 1$, Fig.~\ref{fig:Modes}b) 
to shapes that are ``flat'' ($\rho/r > 1$, Fig.~\ref{fig:Modes}c)
at their contact.
Both conditions are illustrated in Figs.~\ref{fig:Modes}b and~c.
In the contour plot, we present a range of dimensionless force 
$f/(kr)$ that is fairly narrow, from $-0.005$ to $0.02$.
The positive, compressive values are of a range typical
for hard particles; whereas, the negative values could
occur in dry powders when electrostatic and van der Waals 
attractions are active.
As expected, sharp contacts are unstable ($\lambda^{(6)}<0$)
and flat contacts are stable ($\lambda^{(6)}>0$)
for any compressive force $f>0$.
This result, although limited to a simple two-particle system,
is consistent with the widely observed
tendency of granular materials toward stress-induced anisotropy,
in which contacts become predominately flat-to-flat in the direction of
compressive loading \cite{Rothenburg:1993a}.
In regard to uniqueness, 
Type~2 neutral equilibrium
occurs under conditions that produce $\lambda^{(6)}=0$:
either with circular disks ($\rho/r=1$) or with zero-force, 
grazing contacts ($f=0$).
\par
When two \emph{circular} disks are pressed together, they are in
neutral equilibrium and neutral stability, with $\lambda^{(6)}=0$.
\begin{figure}
\centering
\includegraphics[scale=0.90]{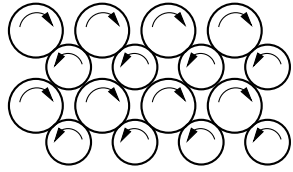}
\caption{A gear-like bifurcation mode in a regular packing when the
rolling stiffness $\mathbf{M}^{pq}=\mathbf{0}$.}
\label{fig:gears}
\end{figure}
For example, a bifurcation of motions is readily available
to the system in Fig.~\ref{fig:gears}:  a synchronized,
gear-like turning
of the disks can be superposed onto any other solution.
This bifurcation would, of course, be inhibited by any genuine rotational
stiffness at the contact, demonstrating that the possible bifurcation mode in
Fig.~\ref{fig:gears} is simply a consequence of the constitutive choice
$\mathbf{M}^{pq}=\mathbf{0}$ in Eq.~(\ref{eq:constitutiveM}).
\subsection{Four-disk system}\label{sec:fourdisks}
\begin{figure}
\centering
\mbox{%
\subfigure[]{\raggedright\includegraphics[scale=0.90]{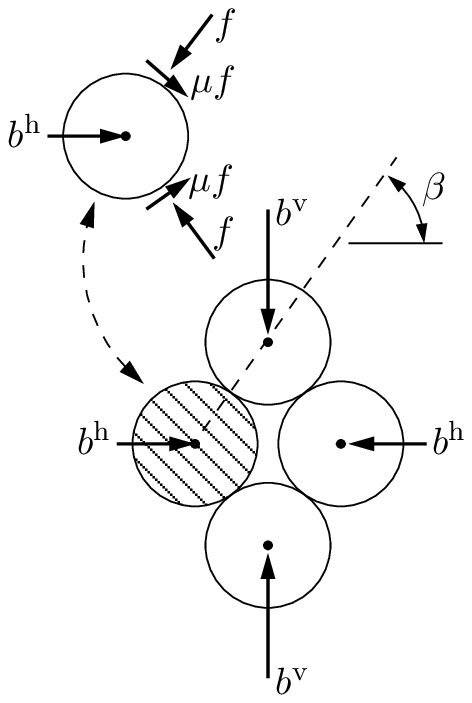}}\ \quad%
\subfigure[]{%
\parbox[b]{1.6in}{%
  \includegraphics[scale=0.60]{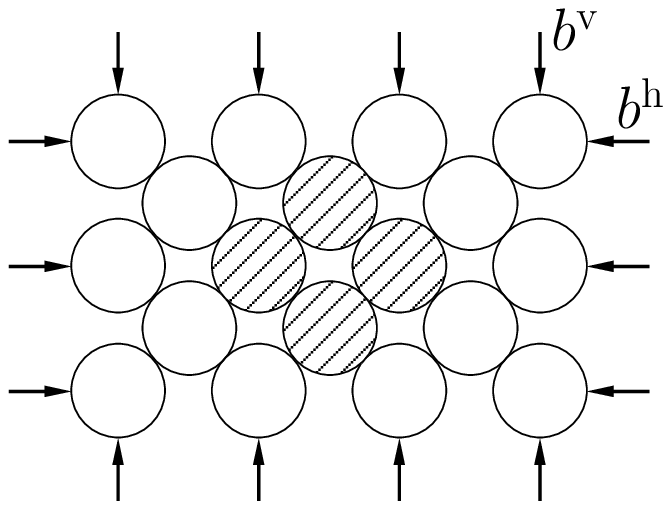}%
  \\[0.3in]\rule{0in}{0.01in}}\quad%
}
\subfigure[]{\includegraphics[scale=0.90]{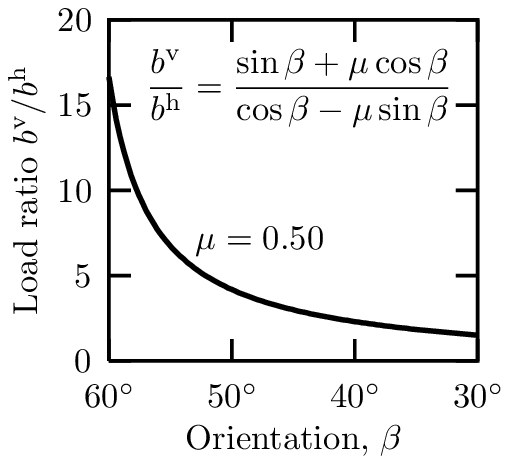}}
}
\caption{Four-disk example.}
\label{fig:fourdisk}
\end{figure}

We now analyze an isolated cluster of four equal-size
disks having four contacts (Fig.~\ref{fig:fourdisk}a), noting that
this cluster might represent the repeating unit of a regular 
2D assembly (Fig.~\ref{fig:fourdisk}b).
We assume that
the four disks have been compressed vertically
while they have expanded horizontally,
so that current opposing pairs of vertical and
horizontal external forces, $b^{\text{v}}$ and
$b^{\text{h}}$, produce a frictional sliding
at all four contacts (Fig.~\ref{fig:fourdisk}a).
The system would soften 
under these loading conditions, as shown by plotting
the force ratio $b^{\text{v}} / b^{\text{h}}$ against the angle $\beta$
(Fig.~\ref{fig:fourdisk}c).
We examine the system at a given angle $\beta$ to determine
the eigenmodes of further (incremental) deformation.
Since all four contacts are known to be sliding
at angle $\beta$ ($M^{\text{s}}=4$), the subsequent motions present
$L=2^{4}=16$ possible combinations 
(i.e. branches) of contact loading or unloading
(sliding or elastic sticking).
Each combination is a separate, $i$th, branch of the cluster
stiffness $\mathbf{H}(\Dut)$.
We must construct the mechanical stiffness 
$[\HH^{\text{m, }i}]$ for each branch
and then add it to the shared geometric stiffness $[\Hg ]$,
which will be the same for all branches.
The sixteen combined stiffnesses $[\HH^{i}]$
are $12\times 12$,
since every 2D particle has three degrees of freedom.
With each loading-unloading combination, we find the twelve
eigenvalues and eigenvectors 
of its matrices $[\HHH^{i}]$ and $[\widehat{\HHHH}^{i}]$
and then determine which of the
eigenvectors are consistent with the presumed combination
of loading and unloading for this branch
(Sections~\ref{sec:uniqueness} and~\ref{sec:softening}).
The question of whether an eigenvector produces a consistent
loading-unloading combination is determined by applying
Eq.~(\ref{eq:flow}) to each of the four contacts.
Appendix~\ref{app:eigen} describes a search algorithm.
\par
Numerical results were developed for the following conditions:
equal normal and tangential contact stiffnesses ($\alpha = 1$),
compressive contact forces that are much smaller than the contact
stiffness ($f/k = 1/1000$),
a friction coefficient $\mu=0.5$, and a particle orientation
$\beta=45^{\circ}$.
We assume that all four contacts are currently sliding
($Q=0$ in Eq.~\ref{eq:yield}),
but allow the possibility that all (or some) contacts cease
slipping during the subsequent motion $\Dut$.
\par
The results show that each of the sixteen stiffness branches
$[\widehat{\HHHH}^{i}]$ has four zero-eigenvalues:
three of these eigenvalues correspond to rigid-body motions;
the fourth corresponds to a gear-like rolling mode, such
as that depicted
in Fig.~\ref{fig:gears}.
Regardless of the branch that is active in a 
loading increment $[\D\mathbf{b}/\D\mathbf{w}]$,
the system has no better than neutral stability
(Eq.~\ref{eq:neutral}), since the gear-like mode
presents a zero-work increment that can be superposed
on any solution.
The sixteen branches $[\widehat{\HHHH}^{i}]$
possess a total of 30 non-zero eigenvalues whose
eigenvectors are consistent with the loading-unloading combination of
their respective branches
(Section~\ref{sec:stability} and Appendix~\ref{app:eigen}).
Twenty-one of these eigenvalues are positive; nine are negative.
The presence of multiple negative eigenvalues indicates
that the cluster is unstable: small changes in the external
forces $b^{\text{v}}$ and $b^{\text{h}}$ can produce large
displacements and a loss of the cluster's capacity to support a
sustained, dead load.
The negative eigenvalues also indicate that even if the displacements
can be controlled, the system will soften along numerous load paths,
such as the one shown in Fig.~\ref{fig:fourdisk}c.
\par
The cluster's instability and its potential for softening have two sources.
Frictional contact sliding is inherently unstable
and can produce softening by means of the cluster's mechanical
stiffness $[\Hm ]$.
The mechanical stiffness is a collection of contact
stiffnesses, and the symmetric part of the 
frictional contact stiffness 
$[\widehat{\mathbf{F}}^{pq}]$ in Eq.~(\ref{eq:ContactF}$_{2}$)
has a negative eigenvalue of $(1-\sqrt{1+\mu^{2}})/2$.
Ba{\u{z}}ant and Cedolin (1991, \S10.7) show that negative 
second-order work is produced in a single-body frictional system
through the release of frictionally
blocked elastic energy, even though the system
is otherwise stable when the displacements are controlled.
We suspect that the softening observed in many granular 
materials is due, in part, to this mechanical origin.
Instability and softening can also originate from the geometric
stiffness $[\Hg]$.
This origin is illustrated in Fig.~\ref{fig:fourdisk}c,
which shows the softening that ensues when the particles
do not rotate and sliding continues on all four contacts.
During such vertical compression, the magnitudes 
of the normal and tangential forces can be maintained constant
(i.e. constant $f$ and $\mu f$ forces in Fig.~\ref{fig:fourdisk}a).
No frictionally blocked elastic energy is released during the softening
shown in Fig~\ref{fig:fourdisk}c.
All of this softening has a geometric origin.
\par
The two examples reveal the importance of including the geometric
stiffness $[\Hg]$ when evaluating stability.
In both examples, instability and softening are attributed
to the influence of $[\Hg]$.
\par
The two examples are readily amenable to analytical or computational
analysis, since the two systems have few particles and only a
few sliding contacts---the number of branches,
$L=2^{M^{\text{s}}}$, is one in the first example and sixteen in the second.
Similar eigenvalue analyses may be impossible for entire
systems of thousands of particles, although the methods in the
examples can be readily applied to clusters within larger systems.
\section{Discussion and Conclusion}
This work provides a conceptual framework for including the 
influence of particle shape on granular stiffness and for
evaluating the potential for instability and softening.  
This approach may be productive in investigating
granular behavior, particularly at large strains.
We foresee three applications:  
(1) as a way of improving current numerical simulation methods 
for granular assemblies, 
(2) as an approach toward understanding granular failure 
and localization, and 
(3) as a means of analyzing and post-processing 
simulation results for understanding granular behavior.
In regard to the first application, 
GEM and DDA simulations methods currently use a 
similar direct stiffness approach to simulate the 
interactions of particles in a granular assembly, 
and these methods could benefit from the full inclusion 
of all stiffness terms of order $(du)^{1}$---terms 
of both mechanical and geometric origin.
\par 
With respect to the second application,
the formulations show that material stiffness depends upon the 
contact stiffnesses and on a complex interaction of the contact 
forces and particle shapes.  
The influence of contact stiffness is embodied
in a mechanical stiffness $[ \Hm ]$, 
and the effects of contact force and particle shape 
are gathered into a geometric stiffness $[\mathbf{H}^{\text{g}}]$.  
The latter stiffness likely has negligible influence at small strains, 
but its effect may become substantial, perhaps dominant, during failure:  
at large strains, the rotation and rolling among nearly rigid particles
become prevalent kinematic mechanisms---conditions in which
the geometric stiffness is most active.
Moreover, the bulk stiffness of
granular materials is small or even negative during failure, 
and the otherwise small geometric stiffness likely becomes a relatively
larger contributor during failure.  
Because the geometric stiffness is proportional to the current, 
accumulated contact forces, our approach might also explain
why many aspects of granular failure are 
influenced by the confining pressure.
The confining pressure is known to influence the strain at peak stress, 
the friction angle at the peak stress,
the dilation rate at the peak stress, 
the strain at which shear bands begin to appear, 
the orientation and thickness of shear bands,
and the rate of softening at post-peak strains 
\cite{Lee:1967a,Desrues:2004a}.  
A comprehensive micro-mechanical explanation is 
currently lacking for such observed behaviors, 
and these phenomena should be examined in the context of the current work.  
The work may also provide a basis for investigating 
local stiffness, stability, and softening within granular regions, 
perhaps within small representative elements of material.  
For example, the shear bands that appear during failure 
are thought to be an ongoing instability in which particle 
chains continually buckle and then reorganize while a specimen 
is being loaded \shortcite{Oda:1998b,Mair:2002a}.  
Just as material behavior at small strains has been
successfully estimated by using simple micro-mechanical models,
the current approach might be useful in investigating material 
behavior and instability within shear bands at larger strains.
\par 
A third application is in post-processing the results of 
DEM simulations to explore local behavior.  
Unlike the GEM and DDA methods, 
the DEM does not use a direct stiffness approach, 
but instead uses an efficient dynamic relaxation algorithm to track 
the interactions of particles while an assembly is being deformed
\cite{Cundall:1979a}.  
Methods have already been proposed for 
extracting the spatial distributions  
of stress and strain from DEM results 
\cite{Bagi:1996a,Satake:2004a}.  
The current work provides a means of quantifying 
local stiffness within granular materials, 
so that questions of instability and softening can be studied 
through DEM simulations:  
the simulations would provide the state of a granular assembly;
whereas, the current methods could be used to explore the
stiffness characteristics in that state.
\par
Finally, we note that most existing simulation 
methods---GEM, DDA, and DEM---are meant to solve 
large boundary value problems that involve 
a discrete, granular region, 
and the success of a simulation is often 
judged by the numerical stability of its algorithm.  
These methods can provide a solution, 
but without determining whether non-unique, 
multiple solutions are possible at any stage of loading.  
The proposed stability and uniqueness criteria provides a 
framework for investigating the
stability and possible bifurcation of solutions during loading.
\section*{Acknowledgement}
Katalin Bagi assisted in the current work through her insightful
discussions.
She presents a parallel derivation of matrix $[\mathbf{H}]$
that compliments the current work \cite{Bagi:2005a}.
\pagebreak
\appendix
{\noindent\\[2ex]\LARGE\bfseries Appendices}%
\section{Notation}\label{sec:notation}
The following symbols are used in this paper:
\nopagebreak
\par
\begin{tabular}{r  @{\hspace{1em}=\hspace{1em}}  p{5.25in}}
$[\mathbf{A}_{1}]$ & statics matrix, particle group,~(\ref{eq:Matrix2}) \\
$[\mathbf{A}_{2}]$ & contact force rotation 
                     matrix, particle group,~(\ref{eq:A3}) \\
$\Bp$ & external force on $p$, Fig.~\ref{fig:particles} \\
$[\mathbf{B}]$ &  kinematics matrix, particle group, (\ref{eq:kinematics}) \\
$[\mathbf{C}]$ & rigid rotation matrix, particle group, (\ref{eq:dtrigid}) \\
$\Delbp$, $\Dbp$ & increment, external force 
          on $p$,~(\ref{eq:equil2}), (\ref{eq:db}), (\ref{eq:ddb})\\
$\Delfpq$, $\Dfpq$, $\DFfpq$ & increment, contact force on $p$ by $q$,
               (\ref{eq:equil2}), (\ref{eq:df}), and~(\ref{eq:delfpq}) \\
$\Delmpq$, $\Dmpq$, $\DMfpq$ & increment, contact moment on $p$ by $q$,
               (\ref{eq:equil2}), (\ref{eq:dm}), and~(\ref{eq:delmpq}) \\
$d\mathbf{n}^{pq}$, $\Dnpq$ & increment, surface normal of 
               $p$ at contact $pq$, (\ref{eq:dn}) and~(\ref{eq:Dnpq})\\
$d\Rpq$, $\Drpq$ & increment, contact radius, 
              (\ref{eq:equil2}), (\ref{eq:dr}), and~(\ref{eq:Drpq})\\
$d\mathbf{u}^{p}$, $\D\mathbf{u}^{p}$ & translation of $p$, 
                        Fig.~\ref{fig:particles}, (\ref{eq:piece}),
                        and~(\ref{eq:ddu})\\
$\Delwp$, $\Dwp$ & increment, external force on
       $p$,~(\ref{eq:equil2}), (\ref{eq:dw}), (\ref{eq:ddb})\\
$d\boldsymbol{\theta}^{p}$, $\D\boldsymbol{\theta}^{p}$ & 
                        rotation of $p$,
                        Fig.~\ref{fig:particles}, (\ref{eq:piece}),
                        and~(\ref{eq:ddu})\\
$d\overline{\boldsymbol{\theta}}^{\text{rigid}}$ &
       rigid rotation, particle group, (\ref{eq:dutrigid})--(\ref{eq:Moore1})\\
$\mathbf{f}^{pq}$ & contact force on $p$ by $q$,
                    (\ref{eq:equil1})\\
$[\mathbf{F}/\mathbf{M}]$ & contact constitutive matrix, particle group,
                            (\ref{eq:constitMatrix})\\
$\mathbf{F}^{pq}$ & contact stiffness tensor, 
                    (\ref{eq:constitutive}) and (\ref{eq:ContactF})\\
$\mathbf{g}^{pq}$ & yield surface normal, contact $pq$,
                    (\ref{eq:G})\\ 
$\mathbf{h}^{pq}$ & sliding direction, contact $pq$, (\ref{eq:Hdirection})\\
$[\mathbf{H}]$ & stiffness matrix, particle group,~(\ref{eq:H}) 
                 and~(\ref{eq:final2}) \\
$[\mathbbm{H}]$, $[\HHHH]$ & modified stiffness matrices, particle group,
                         (\ref{eq:newstiffnesses})\\
$[\widehat{\HHHH}^{i}]$ & symmetric part of $[\HHHH]$\\
$[\Hg ]$ & combined geometric stiffness, particle group,
           (\ref{eq:final1}) and~(\ref{eq:final2})\\
$[\Hgone]$, $[\Hgtwo]$, $[\Hgthree]$ & geometric stiffnesses, particle group,
              (\ref{eq:Matrix2}), (\ref{eq:equilibrium4}), 
              (\ref{eq:A4})\\
$[\Hm]$ & mechanical stiffness matrix, particle group, (\ref{eq:Hm})\\
$[\mathbf{I}]$, $\mathbf{I}$ & identity matrix, Kronecker tensor\\
$k$ & contact stiffness, (\ref{eq:ContactF})\\
$[\mathbf{K}^{p}]$ & surface curvature tensor of $p$ at contact $pq$,
                     (\ref{eq:rolling})\\
$\mathbf{m}^{pq}$ & contact moment on $p$ by $q$,
                    (\ref{eq:equil1})\\
$\mathbf{M}^{pq}$ & contact rotational stiffness tensor, contact $pq$,
                    (\ref{eq:constitutiveM})\\
$\mathbf{n}^{pq}$ & unit normal vector, outward from $p$ toward $q$,
                    (\ref{eq:Drpq})
\end{tabular}
%
%
\par
\begin{tabular}{r  @{\hspace{1em}=\hspace{1em}}  p{5.25in}}
$N$ & number of particles, particle group \\
$[\mathbf{P}^{\text{n--r--r}}]$ & projection onto no-rigid rotation subspace, 
                               particle group, (\ref{eq:Pnrr})\\
$[\mathbf{P}^{\text{r--r}}]$ & projection onto rigid rotation subspace, 
                               particle group, (\ref{eq:Prr})\\
$Q^{pq}$ & contact sliding condition, contact $pq$,
           (\ref{eq:yield})\\
$\mathbf{r}^{pq}$ & particle radial vector, from $\mathbf{x}_{p}$ to 
                    contact $pq$, Fig.~\ref{fig:particles}\\
$S^{pq}$ & contact sliding condition, contact $pq$,
           (\ref{eq:flow})\\
$\mathbf{t}^{pq}$ & unit tangent vector, from $p$ at contact $pq$,
                    (\ref{eq:Drpq})\\
$\Wp$ & external moment on $p$, Fig.~\ref{fig:particles} \\
$\mathbf{x}^{p}$ & position, particle $p$, Fig.~\ref{fig:particles} \\
$\alpha$ & tangential-to-normal contact stiffness ratio,
           (\ref{eq:ContactF})\\
$\beta$ & particle orientation, Fig.~\ref{fig:fourdisk}\\
$\delta s^{pq,\text{n}}$ & normal contact displacement, viewed by $p$,
                           (\ref{eq:Drpq}) and~(\ref{eq:dsn})\\
$\delta s^{pq,\text{t}}$ & tangential contact displacement, viewed by $p$,
                           (\ref{eq:Drpq}) and~(\ref{eq:rolling})\\
\end{tabular}
\section{Derivations of Eqs.~\ref{eq:equil3}, 
\ref{eq:equil4}, and \ref{eq:Dfpq}} \label{app:derive}
Equation~(\ref{eq:equil3}) is derived from Eq.~(\ref{eq:equil2}$_1$)
as follows.  We substitute Eqs.~(\ref{eq:df}) and~(\ref{eq:db})
into Eq.~(\ref{eq:equil2}$_1$),
\begin{equation}
-\Sum\Dfpq - \Dtp\times\Sum\Fpq = \Dbp + \Dtp\times\Bp \;,
\end{equation}
and apply equilibrium Eq.~(\ref{eq:equil1}$_1$) to arrive at 
Eq.~(\ref{eq:equil3}):
\begin{equation} \tag{\ref{eq:equil3}}
-\Sum\Dfpq = \Dbp \;.
\end{equation}
\par
Equation~(\ref{eq:equil4}) is derived from Eq.~(\ref{eq:equil2}$_2$)
by substituting Eqs.~(\ref{eq:dr}), (\ref{eq:df}), and~(\ref{eq:dw}):
\begin{multline}
-\Sum\left(\Drpq\times\Fpq + \Rpq\times\Dfpq + \Dmpq \right)
  \\- \Sum\left[
    \left(\Dtp\times\Rpq\right)\times\Fpq +
    \Rpq\times\left(\Dtp\times\Fpq\right) + \Dtp\times\Mpq
    \right] \\= \Dwp + \Dtp\times\Wp 
\end{multline}
The vector triple product satisfies the identity
$(\mathbf{a}\times\mathbf{b})\times\mathbf{c}=
-\mathbf{b}\times(\mathbf{a}\times\mathbf{c})
+\mathbf{a}\times(\mathbf{b}\times\mathbf{c})$, so that
\begin{multline}
  -\Sum\left(\Drpq\times\Fpq + \Rpq\times\Dfpq + \Dmpq\right)
  \\
  - \Dtp\times\Sum\left[
      \left( \Rpq \times\Fpq\right) + \Mpq \right] =
  \Dwp + \Dtp\times\Wp \;,
\end{multline}
and applying Eq.~(\ref{eq:equil1}$_2$),
\begin{equation}
-\Sum\left(\Drpq\times\Fpq + \Rpq\times\Dfpq + \Dmpq\right) = \Dwp \;.
\tag{\ref{eq:equil4}}
\end{equation}
\par
Equation~(\ref{eq:Dfpq}) is derived from Eq.~(\ref{eq:delfpq}) as follows.  
We substitute the definition~(\ref{eq:dtdef}) of $\Dtdef$
into Eq.~(\ref{eq:delfpq}):
\begin{equation} \label{eq:dfeval}
\begin{split}
\Delfpq = \;&\DFfpq + \Fpq\times\left( d\mathbf{n}^{pq}\times\Npq\right)\\
   &-\left( \Dtp\cdot\Npq \right)\Fpq\times\Npq
    - (1/2)\left(\Dtdef\cdot\Npq\right)\Fpq\times\Npq \;.
\end{split}
\end{equation}
and then substitute Eq.~(\ref{eq:dn}),
\begin{equation}\label{eq:dfinter}
\begin{split}
\Delfpq = &\DFfpq + \Fpq\times(\Dnpq\times\Npq)
+ \Fpq\times[(\Dtp\times\Npq)\times\Npq]\\
&-(\Dtp\cdot\Npq)\Fpq\times\Npq
-(1/2)(\Dtdef\cdot\Npq)\Fpq\times\Npq
\end{split}
\end{equation}
Taking the third term on the right, we apply the identity
$\mathbf{a}\times(\mathbf{b}\times\mathbf{c})
=(\mathbf{c}\cdot\mathbf{a})\mathbf{b}
-(\mathbf{b}\cdot\mathbf{a})\mathbf{c}$
and the aforementioned vector triple product identity,
\begin{equation}
\Fpq\times[(\Dtp\times\Npq)\times\Npq]
=
\Dtp\times\Fpq + (\Dtp\cdot\Npq)\Fpq\times\Npq
\end{equation}
This relation and Eq.~(\ref{eq:df}) are substituted in
Eq.~(\ref{eq:dfinter}) to find Eq.~(\ref{eq:Dfpq}).
\section{Derivations of two-particle example,
Section~\ref{sec:twoparticle}} \label{app:example}
In this appendix, the stiffness matrix is derived for the simple two-particle
system of Section~\ref{sec:example}.  
The particle arrangement is shown in Fig.~\ref{fig:Example} and
the related data is summarized in Table~\ref{table:Example}.
\renewcommand{\arraystretch}{1.4}
\begin{table}
\centering
\caption{Data for the two-particle cluster in Fig.~\ref{fig:Example}.}
\begin{tabular}{ll}
\hline
Object & Value \\
\hline
$\Fpq = -\Fqp = -\mathbf{b}^{p} = \mathbf{b}^{q}$ & $[-f,\;0]^{\text{T}}$ \\
$\Mpq = -\Mqp = \mathbf{w}^{p}=\mathbf{w}^{q}$ & 0 \\
$\Npq = -\Nqp$ & $[1,\; 0]^{\text{T}}$\\
$\Rpq$, $\Rqp$ & $[r^{p},\; 0]^{\text{T}}$, $[-r^{q},\; 0]^{\text{T}}$ \\
$[\Kp]$, $[\Kq]$ & 
  $\left[\begin{smallmatrix}0&0\\0&-1/\rho^{p} \end{smallmatrix}\right]$,
  $\left[\begin{smallmatrix}0&0\\0&-1/\rho^{q} \end{smallmatrix}\right]$ \\
$[\Kp + \Kq]^{-1}$ &
  $\left[\begin{smallmatrix}0&0\\0&-\rho^{p}\rho^{q}/(\Rhopq)
         \end{smallmatrix}\right]$ \\
$[\mathbf{F}^{pq}] = -[\mathbf{F}^{qp}]$ &
  $\left[\begin{smallmatrix} k&0\\0&\alpha k
         \end{smallmatrix}\right]$ \\
$[\mathbf{M}^{pq}] = -[\mathbf{M}^{qp}]$ &
  $\left[\begin{smallmatrix} 0&0\\0&0
         \end{smallmatrix}\right]$ \\
\hline
\end{tabular}
\label{table:Example}
\end{table}
\par
The two geometric stiffnesses $[\Hgone ]$ and
$[\mathbf{H}^{\text{g--2}}]$ depend upon the movements
$\Dst\Tpq$ and $\delta s^{qp\text{, t}}\Tqp$ in
Eqs.~(\ref{eq:Drpq})--(\ref{eq:dtdef}).
For the data in Table~\ref{table:Example},
\begin{align}
\label{eq:tpq}
\Dst\Tpq &=
- \left( \frac{-\rho^{p} \rho^{q}}{\Rhopq}\right)
\left[
\left( d\theta^{q}_{3} - d\theta^{p}_{3}\right)
- ( \frac{-1}{\rho^{q}})
\left( du^{q}_{2}-du^{p}_{2}-d\theta^{p}_{3}r^{p}-d\theta^{q}_{3}r^{q}\right)
\right] \Etwo
\\
\label{eq:tqp}
\delta s^{qp\text{, t}}\Tqp &=
- \left( \frac{-\rho^{p} \rho^{q}}{\Rhopq}\right)
\left[
\left( d\theta^{q}_{3} - d\theta^{p}_{3}\right)
+ ( \frac{-1}{\rho^{p}})
\left( du^{q}_{2}-du^{p}_{2}-d\theta^{p}_{3}r^{p}-d\theta^{q}_{3}r^{q}\right) \right] \Etwo
\end{align}
Stiffness $[\Hgone ]$ is defined in
Eqs.~(\ref{eq:equil4}) and~(\ref{eq:Hg1}) as
\begin{equation}\label{eq:examplehgone}
\LB \Hgone \RB \DUa
=
\left[ \begin{MAT}(r)[2pt]{c}
\Zero\\:
-\Drpq \times \Fpq \\:
\Zero\\:
-\Drqp \times \Fqp \\
\end{MAT} \right]
\end{equation}
where the rows have been rearranged to produce forces in the order
$[d\Bp, d\Wp, d\Bq, d\Wq]^{\text{T}}$.
Because the indentations $\Dsn\Npq$ in Eqs.~(\ref{eq:Drpq}) and~(\ref{eq:dsn})
are aligned with the force $\Fpq$, only the tangential rolling 
motions in Eqs.~(\ref{eq:tpq}) and~(\ref{eq:tqp}) contribute
to $[\Hgone ]$, so that the right side of Eq.~(\ref{eq:examplehgone}) is
\begin{equation}
\LB \Hgone \RB \DUa
=
\left[ \begin{MAT}[2.5pt]{ccc:rcc}
0 & 0 & 0 & \ 0 & 0 & 0 \\
0 & 0 & 0 & 0 & 0 & 0 \\
0 & \frac{f\rho^p}{\Rhopq} & \frac{f\rho^p(r^p + \rho^q)}{\Rhopq} & 
0 & \frac{-f\rho^p}{\Rhopq} & \frac{f\rho^p(r^q - \rho^q)}{\Rhopq} \\:
0 & 0 & 0 & 0 & 0 & 0 \\
0 & 0 & 0 & 0 & 0 & 0 \\
0 & \frac{f\rho^q}{\Rhopq} & \frac{f\rho^q(r^p - \rho^p)}{\Rhopq} &
0 & \frac{-f\rho^q}{\Rhopq} & \frac{f\rho^q(r^q + \rho^p)}{\Rhopq} \\
\end{MAT} \right]
\DUb
\end{equation}
The four quadrants in this equation correspond to the
submatrices
$[\mathbf{H}^{\text{g--1, }pp}]$,
$[\mathbf{H}^{\text{g--1, }pq}]$,
$[\mathbf{H}^{\text{g--1, }qp}]$, and
$[\mathbf{H}^{\text{g--1, }qq}]$ of Eq.~(\ref{eq:piece}).
\par
The second geometric stiffness $[\Hgtwo]$ is defined in
Eq.~(\ref{eq:equilibrium4}) as the product
$-[\mathbf{A}_{1}][\mathbf{A}_{2}]$.
The statics matrix $[\mathbf{A}_{1}]$ is
\begin{equation}\label{eq:Aonetwoparticle}
\LB \mathbf{A}_{1} \RB 
\left[ \begin{MAT}(r)[2pt]{c}
\Dfpq\\: \Dmpq\\: \Dfqp\\: \Dmqp\\
\end{MAT} \right]
=
\left[ \begin{MAT}(b)[2.5pt]{ccc:ccc}
1 & 0 & 0 & 0 & 0 & 0 \\
0 & 1 & 0 & 0 & 0 & 0 \\
0 & r^p & 1 & 0 & 0 & 0 \\:
0 & 0 & 0 & 1 & 0 & 0 \\
0 & 0 & 0 & 0 & 1 & 0 \\
0 & 0 & 0 & 0 & -r^q & 1 \\
\end{MAT} \right]
\left[ \begin{MAT}(r)[2pt]{c}
\delta f^{pq}_{1}\\ \delta f^{pq}_{2}\\ \delta m^{pq}_{3}\\:
\delta f^{qp}_{1}\\ \delta f^{qp}_{2}\\ \delta m^{qp}_{3}\\
\end{MAT} \right] \;.
\end{equation}
Matrix $[\mathbf{A}_{2}]$ is defined through Eqs.~(\ref{eq:Dnpq}) 
and~(\ref{eq:A3}),
with
\begin{align}
\Dnpq &= -\Kp\cdot(\Dst\Tpq) = (1/\rho^p) \Dst\Tpq \\
\Dnqp &= -\Kq\cdot(\delta s^{qp\text{, t}}\Tqp) = 
           (1/\rho^q) \delta s^{qp\text{, t}}\Tqp \;,
\end{align}
which is combined with Eqs.~(\ref{eq:tpq}) and~(\ref{eq:tqp}) to find
\begin{equation}
\LB \mathbf{A}_{2} \RB   \DUa
=
\left[ \begin{MAT}[2.5pt]{ccc:rcc}
0 & 0 & 0 & \ 0 & 0 & 0 \\
0 & \frac{f}{\Rhopq} & \frac{f(r^p + \rho^q)}{\Rhopq} &
0 & \frac{-f}{\Rhopq} & \frac{f(r^q - \rho^q)}{\Rhopq}\\
0 & 0 & 0 & 0 & 0 & 0 \\:
0 & 0 & 0 & 0 & 0 & 0 \\
0 & \frac{-f}{\Rhopq} & \frac{-f(r^p - \rho^p)}{\Rhopq} &
0 & \frac{f}{\Rhopq} & \frac{-f(r^q + \rho^p)}{\Rhopq}\\
0 & 0 & 0 & 0 & 0 & 0 \\
\end{MAT} \right] \;,
\DUb
\end{equation}
so that the product $[ \Hgtwo ]=-[\mathbf{A}_{1}] [\mathbf{A}_{2}]$
is
\begin{equation}
\LB \Hgtwo \RB \DUa 
= 
\frac{f}{\Rhopq}
\left[ \begin{MAT}[2.5pt]{ccc:rcc}
0 & 0 & 0 & \ 0 & 0 & 0 \\
0 & -1 & -(r^p + \Rhoq) & 0 & 1 & \Rhoq - r^q \\
0 & -r^p & -r^p(r^p+\Rhoq) & 0 & r^p &
                                                 -r^p(r^q - \Rhoq)\\:
0 & 0 & 0 & 0 & 0 & 0 \\
0 & 1 & r^p - \Rhop & 0 & -1 & r^q + \Rhop \\
0 & -r^q & -r^q(r^p - \Rhop) & 0 & r^q & -r^q(r^q+\Rhop)\\
\end{MAT} \right]
\DUb \;.
\end{equation}
The geometric stiffness $[\Hgthree ]$ in Eq.~(\ref{eq:Hg3})
receives two contributions of the form
$-d\boldsymbol{\theta}\times \mathbf{f}$:
a contribution $f\,d\theta_{3}^{p}\boldsymbol{e}_{2}$ for the ``$pq$''
contact and $-f\,d\theta_{3}^{q}\boldsymbol{e}_{2}$ for the ``$qp$''
contact.
The matrix $[\Hgthree ]$ is
\begin{equation}
\LB \Hgthree \RB   \DUa =
\left[ \begin{MAT}[4.5pt]{ccc:rcc}
0 & 0 & 0 & \ 0 & 0 & 0\\
0 & 0 & f & 0 & 0 & 0\\
0 & 0 & 0 & 0 & 0 & 0\\:
0 & 0 & 0 & 0 & 0 & 0\\
0 & 0 & 0 & 0 & 0 & -f\\
0 & 0 & 0 & 0 & 0 & 0\\
\end{MAT} \right]
\DUb \;.
\end{equation}
The mechanical stiffness $[ \Hm ]$ is defined
in Eq.~(\ref{eq:Hm}), with
\begin{equation}
\left[ \begin{MAT}{c}
\delta u_{1}^{pq\text{, def}}\\
\delta u_{2}^{pq\text{, def}}\\
\delta \theta_{3}^{pq\text{, def}}\\
\end{MAT} \right]
=
\LB \mathbf{B} \RB \DUa =
\left[ \begin{MAT}[4.5pt]{ccc:rcc}
-1 &  0 &  0 & 1 & 0 & 0\\
 0 & -1 &  r^{p} & 0 & 1 & -r^{q}\\
 0 &  0 & -1 & 0 & 0 & -1\\
\end{MAT} \right]
\DUb \;,
\end{equation}
and matrix $[\mathbf{F} / \mathbf{M}]$ defined by 
Eqs.~(\ref{eq:constitMatrix}) and~(\ref{eq:simple}):
\begin{equation}
\left[ \begin{MAT}{ccc:ccc}
\mathfrak{d}f_{1}^{pq} &
\mathfrak{d}f_{2}^{pq} &
\mathfrak{d}m_{3}^{pq} &
\mathfrak{d}f_{1}^{qp} &
\mathfrak{d}f_{2}^{qp} &
\mathfrak{d}m_{3}^{qp}\\
\end{MAT} \right]^{\text{T}}
=
\left[ \begin{MAT}{ccc}
k & 0 & 0\\
0 & \alpha k & 0\\
0 & 0 & 0\\:
-k & 0 & 0\\
0 & -\alpha k & 0\\
0 & 0 & 0\\
\end{MAT} \right]
\left[ \begin{MAT}{c}
\delta u_{1}^{pq\text{, def}}\\
\delta u_{2}^{pq\text{, def}}\\
\delta \theta_{3}^{pq\text{, def}}\\
\end{MAT} \right] \;.
\end{equation}
When combined with $[\mathbf{A}_{1}]$ in Eq.~(\ref{eq:Aonetwoparticle}),
the result
$[ \Hm ]=-[\mathbf{A}_{1}][\mathbf{F} / \mathbf{M}]
[\mathbf{B}]$ is given in Eq.~(\ref{eq:exampleH2}).
The rotation vector $[\mathbf{C}]$, defined in Eq.~(\ref{eq:dtrigid}) is
\begin{equation} \label{eq:exampleC}
\LB \mathbf{C} \RB^{\text{T}} =
\LB 0, -(r^{p} + r^{q})/2, 1, 0,(r^{p} + r^{q})/2,1 \RB \;.
\end{equation}
\section{Algorithm for finding consistent eigenmodes}\label{app:eigen}
An algorithm is required for organizing the eigenvectors
of each branch of $[\HHH^{i}]$ or $[\widehat{\HHHH}^{i}]$ 
and finding the eigenvectors
that are consistent with the loading conditions of their branch.
We assume the contact behavior presented in Section~\ref{sec:epstiffness}.
For each branch of $[\HHH^{i}]$ and $[\widehat{\HHHH}^{i}]$,
an $M$-element 
\emph{mask} vector is ascribed to the particular combination
of contact loading~($+1$) and unloading~($-1$) of that branch.
In the four-contact system of Section~\ref{sec:fourdisks},
all four contacts were assumed to be previously sliding, 
so that sixteen branches must be investigated.
The mask
$[1, -1, -1, -1]$ would designate the 
branch of continued incremental loading (slip)
for the first contact but unloading (elastic stick) in
the other three contacts.
Sixteen combinations of 1's and $-1$'s are possible in 
this four-contact example.
If instead, one of the four contacts has not yet begun to slip 
(e.g., the current contact yield condition,
$Q=0$, in Eq.~\ref{eq:yield} is false),
then only eight branches are available, and
a zero is placed in the mask for the one non-yielding contact, 
regardless of the branch.
After finding the eigenvector for a particular eigenvalue,
a \emph{test vector} is created for the eigenvector:
the test in Eq.~(\ref{eq:flow}) is applied to each contact,
with a 1 (true, $S>0$), $-1$ (false $S<0$),
or 0 (neutral, $S=0$) placed into each contact's position
in the test vector.
If the mask vector matches an
eigenvector's test vector, then the eigenvector
is consistent with its loading-unloading assumptions.
To this end, we find the element-wise product of the mask and
test vectors.  
If each product is 0, then the eigenvector (or the negative of the
eigenvector) is a consistent solution;
if each product is 1 or 0, then the eigenvector
is a consistent solution; 
if each product is $-1$ or 0, then the negative
of the eigenvector is a consistent solution; 
but if any two elements of the product differ in sign, then
the eigenvector is not a consistent solution and must be discarded.
%
%

\end{document}